\documentclass[twocolumn]{aastex6}
\usepackage{amsmath,ulem}
\usepackage{color}
\usepackage{natbib}
\usepackage{graphicx}%, geometry}
\usepackage{epstopdf, xcolor,soul}
\usepackage{verbatim}
\usepackage{comment}
\usepackage{mathtools}
\newcommand{\comments}[1]{} %usage: \comments{}
\newcommand\T{\rule{0pt}{2.6ex}}         % Top strut
\newcommand\B{\rule[-1.2ex]{0pt}{0pt}} % Bottom strut

\shorttitle{Evolution of galactic angular momentum. I.}
\shortauthors{Soko{\l}owska et al.}

\watermark{}

\begin{document}

\title{Galactic angular momentum in cosmological zoom-in simulations. \\ I. {Disk and bulge components and the} galaxy--halo connection}

\author{Aleksandra Soko{\l}owska\altaffilmark{1,5}, Pedro~R. Capelo\altaffilmark{1}, S.~Michael Fall\altaffilmark{2}, Lucio Mayer\altaffilmark{1,5}, Sijing Shen\altaffilmark{3}, \& Silvia Bonoli\altaffilmark{4}}

\email{alexs@physik.uzh.ch}

\altaffiltext{1}{Center for Theoretical Astrophysics and Cosmology, Institute for Computational Science, University of Zurich, \\Winterthurerstrasse 190, CH-8057 Z\"{u}rich, Switzerland}
\altaffiltext{2}{Space Telescope Science Institute, 3700 San Martin Drive, Baltimore, MD 21218, USA}
\altaffiltext{3}{Kavli Institute for Cosmology, University of Cambridge, Madingley Road, Cambridge CB3 0HA, UK}
\altaffiltext{4}{Centro de Estudios de F\'{i}sica del Cosmos de Arag\'{o}n, Plaza San Ju\'{a}n 1, Planta 2, E-44001 Teruel, Spain}
\altaffiltext{5}{Kavli Institute for Theoretical Physics, Kohn Hall, University of California, Santa Barbara, CA 93106, USA}

%%%%%%%%%
%ABSTRACT
%%%%%%%%%

\begin{abstract}
We investigate the angular momentum evolution of four disk galaxies residing in Milky Way-sized halos formed in cosmological zoom-in simulations with {various} sub-grid physics and merging histories. We decompose these galaxies kinematically and photometrically, into their disk and bulge components. The simulated galaxies and their components lie on the observed sequences in the $j_*$--$M_*$ diagram relating the specific angular momentum and mass of the stellar component. {We find that galaxies in low-density environments follow the relation $j_* \propto M_*^{\alpha}$ past major mergers, with $\alpha \sim 0.6$ in the case of strong feedback, when bulge-to-disk ratios are relatively constant, and $\alpha \sim 1.4$ in the other cases, when secular processes operate on shorter timescales. We compute the retention factors (i.e. the ratio of the specific angular momenta of stars and dark matter) for both disks and bulges and show that they vary relatively slowly after averaging over numerous but brief fluctuations. For disks, the retention factors are usually close to unity, while for bulges, they are a few  times smaller. Our simulations therefore indicate that galaxies and their halos grow in a quasi-homologous way.}
\end{abstract}

%%%%%%%%%
%INTRODUCTION
%%%%%%%%%

\section{Introduction}\label{sec:introduction}

The mass $M$ and angular momentum $J$ are two of the most basic properties of galaxies.  For many purposes, it is more convenient and physically meaningful to describe galaxies in terms of their mass $M$ and specific angular momentum $j \equiv J/M$ (for the stellar parts of galaxies, we denote these quantities by {$M_*$ and $j_*$}).  Galaxies of the same disk-to-{bulge} ratio or morphological type at redshift $z = 0$ obey scaling relations of the form $j_* \propto M_*^{\alpha}$, with $\alpha \sim 2/3$ \citep{Fall:1983,Romanowsky:2012,Fall:2013}. In a plot of $\log j_*$ against $\log M_*$, disk-dominated {(Sc, Sd) and bulge}-dominated (E) galaxies lie along roughly parallel sequences of slope $\sim$2/3 separated by a factor of $\sim$5 in $j_*$ at each $M_*$. Galaxies of intermediate types (Sb, Sa, and S0) populate the region between these sequences.

The observed sequence of disk-dominated galaxies at $z = 0$ in the $j_*$--$M_*$ diagram is close to the predictions of a simple analytical model in which {galactic disks have the same specific angular momenta as their dark-matter halos} \citep{Fall:1980,Fall:1983,Mo:1998}. Disk-dominated galaxies at high redshift ($0.2 < z < 3$) also appear to obey this simple model \citep{Burkert:2016,Contini:2016}. Moreover, the sizes of galactic disks, another reflection of their angular momenta, are consistent with this model, both at $z = 0$ \citep{Kravtsov:2013} and at $0 < z < 3$ \citep{Huang:2016}.

Over the past two decades, there have been many attempts to reproduce the observed $j_*$--$M_*$ sequences in hydrodynamical simulations of galaxy formation. Until recently, most of these simulations produced galaxies that lay closer to the {bulge}-dominated sequence than to the disk-dominated sequence \citep[e.g][]{Navarro:1997overCL, Weil:1998, Abadi:2003, MUGS:2010}. The failure of the early simulations to reproduce the observed disk-dominated sequence has been called the {``angular momentum problem.''} It is another manifestation or close relative of the {``over-cooling problem.''}

{The situation has changed dramatically in the past few years as a result of greater computing power, better numerical techniques, and the inclusion of more realistic physical processes in the simulations. Feedback -- the injection of momentum and/or energy into the interstellar and/or circumgalactic media (ISM and CGM) -- appears to be crucial \citep{Okamoto:2005, Governato:2007, Scannapieco:2008, Zavala:2008}. At the same time, increased mass and spatial resolution have reduced numerical artefacts responsible for spurious angular momentum losses in galactic disks \citep{Okamoto:2003,Kaufmann:2007, Stinson:2010}.  Some of the simulations are now capable of producing respectable galactic disks \citep[e.g.][]{Okamoto:2005, Governato:2007, Scannapieco:2008, Guedes:2011aa, Agertz:2011, Aumer:2013, Marinacci:2014,Roskar:2014, Murante:2015, Colin:2016, Lagos:2016}. The most recent generation of large-volume hydrodynamical simulations of galaxy formation \citep[e.g. Illustris and Eagle;][]{Vogelsberger:2014, Schaye:2015} have also succeeded, at least approximately, in reproducing the observed sequences of disk-dominated, bulge-dominated, and intermediate-type galaxies \citep{Genel:2015,Pedrosa:2015,Teklu:2015,Zavala:2016}.}

{The evolution of the angular momentum of galaxies and their disk and bulge components, relative to their formation history, has been studied in greater detail by \citet{Scannapieco:2009} and \citet{Sales:2012}. Some physical processes cause losses in specific angular momentum \citep[e.g. galaxy mergers;][]{Jesseit:2009,Capelo:2016}, while others cause gains \citep[e.g. galactic fountains;][]{Ubler:2014}. However, the observed $j_*$--$M_*$ diagrams for disk-dominated galaxies clearly show that those gains and losses mostly cancel out, leading to an apparent, if not strict, conservation of specific angular momentum \citep{Romanowsky:2012,Fall:2013}.}

The large-volume hydrodynamical simulations, with typical dimensions $\sim$100~Mpc, produce many thousands of galaxies, more than enough to define the $j_*$--$M_*$ relations over wide ranges of mass and disk-to-bulge ratio. The price paid for these large volumes and galaxy populations, however, is relatively low mass and spatial resolution, typically $\sim$$10^7$~M$_{\odot}$ and $\sim$kpc. Many of the most important physical processes, particularly those involving transport of radiation, mass, and momentum or energy, are affected strongly by the structure of the ISM and CGM on much smaller scales. To take only one example among many, the formation of star clusters and their resulting feedback occurs in some of the densest parts of the ISM, the so-called clumps, with typical masses $\sim$$10^2$--$10^6$~M$_{\odot}$ and dimensions $\sim$pc.

These complicated, and only partially understood, small-scale processes are dealt with in the hydrodynamical simulations of galaxy formation by approximate sub-grid prescriptions rather than by direct solution of the relevant dynamical equations.  Given the large mismatch of scales, the use of sub-grid modules is likely to be necessary in this field for the foreseeable future.  Thus, it is important to analyze simulations with different resolution, numerical techniques, and sub-grid prescriptions, especially for star formation and feedback by both young stars and active galactic nuclei (AGN), to determine which results depend sensitively on these features and which are robust.  In this respect, high-resolution zoom-in simulations of the formation and evolution of individual galaxies are a valuable complement to the large-volume simulations of galaxy populations.  In particular, they allow for a more detailed and reliable study of the processes causing gains and losses of specific angular momentum. Recently, zoom-in simulations have been able to capture even subtle internal dynamical processes occurring in disks, from non-axisymmetric instabilities such as bars to radial migration of stars, and their interplay with other galaxy properties such as the age of stellar populations (\citealt{Brook:2011,Bird:2013aa,Guedes:2011aa,Guedes:2013, Gabor:2013, Stinson:2013Bovy, Bonoli:2016, Spinoso:2016}). {Thus, the latest generation of zoom-in simulations appears to capture many of the salient features of galaxy formation and evolution.}

In this and a companion paper (hereafter papers I and II), we report on a study of the angular momentum in four high-resolution zoom-in simulations of galaxy formation and evolution.  The focus of paper I is on the evolution of the stellar components -- the disks and {bulge}s -- of these galaxies in the $j_*$--$M_*$ diagram from the beginning of the simulations at redshift $z \sim 100$ all the way to the end at $z = 0$. The focus of paper II is on the evolution of the inflowing, outflowing, and circulating gas, in different ranges of density and temperature, how it gains and loses specific angular momentum, and how this accounts for the evolution of the stellar components. Running the simulations all the way to $z = 0$ is crucial, because their behavior changes in important ways at $z \sim 1$. Our study is similar in spirit to the recent analyses of high-resolution zoom-in simulations by \citet{Argo:2014}, \citet{Danovich:2015}, and \citet{Agertz:2016}.

The plan for the remainder of this paper is the following. In Section~\ref{sec:simulations}, we describe our simulations. In Section~\ref{sec:bulge-disk-decomposition}, we decompose the galaxies into disks and {bulge}s. In Section~\ref{sec:j-M-diagrams}, we plot galaxies and their components in the $j$--$M$ diagram at $z = 0$ and at higher redshifts. In Section~\ref{sec:dm-baryon-connection}, we study the relation between the specific angular momentum of galaxies and their dark halos. Finally, in Section~\ref{sec:conclusions}, we discuss and summarize our conclusions. The adopted cosmological parameters in all four simulations are $\Omega_{\rm M} = 0.24$, $\Omega_{\Lambda} = 1 - \Omega_{\rm M}$, $\Omega_{\rm b} = 0.042$, $H_0 = 73$~km~s$^{-1}$~Mpc$^{-1}$, $n_{\rm s} = 0.96$, and $\sigma_8 = 0.76$, based on the first three years of data from the Wilkinson Microwave Anisotropy Probe \citep{Spergel:2007}.

\section{Simulations}\label{sec:simulations}

{We analyze four high-resolution cosmological zoom-in simulations of Milky Way-sized galaxies. The simulations were} performed with the tree-smoothed particle hydrodynamics (SPH) code {\sc gasoline} \citep{Wadsley:aa} with mass resolution $m_{\rm dm} \simeq 9.8 \times 10^4$~M$_{\odot}$ and $m_{\rm SPH} \simeq 2 \times 10^4$~M$_{\odot}$, and spatial resolution $\simeq 120$~{pc. The} zoom-in technique \citep[][]{Katz_White_1993} is well established numerically after more than two decades from its introduction, but care has to be taken in building the initial conditions to avoid numerical artifacts which could affect the dynamics. For the simulations presented here, the original periodic low-resolution box from which the initial conditions were subsequently refined is much larger than the Lagrangian subvolume that was selected for the refinement. The large-scale box has indeed a size of 90 Mpc as opposed to about 1~Mpc for the Lagrangian high-resolution subvolume at $z = 0$ (different for Venus, being 60~Mpc and 0.2~Mpc, respectively). The total number of particles in the box of Eris is 53 million (including 13 million of gas), whereas the box of Venus has 170 million particles (16 million of gas).

While the base box is larger than in other published zoom-in simulations, as explained in \citet{Katz_White_1993}, \citet{Mayer_et_al_2008}, and \citet{Governato:2004}, choosing a large enough box for the coarsely resolved region is important because lack of large-scale power may bias the angular momentum of collapsing halos. In building the initial conditions we also checked that the spin parameter of the selected halo remains essentially unchanged as we introduce successive refinements.

\tabletypesize{\footnotesize}
\begin{deluxetable}{c|c|c|c|c|c|c|c}
\tablecaption{Input parameters of the runs}
\tablenum{1}
\tablehead{\colhead{Run} & \colhead{UVB} & \colhead{IMF} & \colhead{$n_{\rm SF}$} & \colhead{$\epsilon_{\rm SN}$} & \colhead{MC} & \colhead{AGN} & \colhead{IC}}
\startdata
Eris		& HM96 & K93 & 5		& 0.8 & low-T & no  & Q\\
Venus	& HM96 & K93 & 5		& 0.8 & low-T & no  & A\\
EBH		& HM96 & K93 & 5		& 0.8 & low-T & yes & Q\\
E2k		& HM12 & K01 & 100	& 1.0 & all-T   & no  & Q\\
\enddata
\tablecomments{UVB -- UV background (HM96: \citealt{Haardt:1996}, HM12: \citealt{Haardt:2012aa}), IMF -- initial mass function (K93: \citealt{Kroupa:1993aa}, K01: \citealt{Kroupa:2001aa}), $n_{\rm SF}$ -- star formation density threshold, $\epsilon_{\rm SN}$ -- SN efficiency parameter, MC -- metal cooling, AGN -- AGN feedback, and IC -- initial conditions (Q: quiet merger history, A: active merger history).}
\label{tab:tab0}
\end{deluxetable}

One of the runs, Eris, has been shown to be extremely successful in recovering various properties of late-type spirals such as the Milky Way \citep{Guedes:2011aa}. The other runs comprise two which stem from the same initial conditions but have different sub-grid models, and a fourth one with different initial conditions. They are, in order: Eris2k (hereafter E2k, described in more detail in \citealt{Sokolowska:2016}) for which sub-grid parameters were tuned to yield a stronger effect of supernova (SN) feedback to lower star formation rates at high redshift; ErisBH (hereafter EBH) being a replica of Eris that includes AGN feedback and yields final correlations between galaxy properties and the mass of the central supermassive black hole that are in good agreement with those of late-type spirals \citep{Bonoli:2016}; and Venus, with the same sub-grid physics as the original Eris but different initial conditions, chosen to have an active merging history down to low redshift in contrast with the quiet merging history of the other three runs but also a nearly identical final virial halo mass ($\sim 8 \times 10^{11}$~M$_{\odot}$) and halo spin parameter ($\lambda \sim 0.03$). Some important simulation parameters of all runs, including the choice of the UV background and aspects of the sub-grid physics, are listed in Table~\ref{tab:tab0} and discussed below.

All runs include radiative and Compton cooling. However, in Eris, EBH, and Venus{,} gas cooling is computed for a simple mixture of H and He via non-equilibrium cooling rates in the presence of the ionizing cosmic ultraviolet {(UV) background \citep{Haardt:1996,Wadsley:aa}}. Additionally, gas of $T<10^4$~K cools through fine structure and metastable lines of C, N, O, Fe, S, and Si \citep{Bromm:2001aa, Mashchenko:2007aa}. In E2k{,} we instead account for metal-line cooling at all temperatures, employing tabulated rates computed with the code {\sc cloudy} \citep{cloudy}, which assumes that metals are in ionization equilibrium \citep{Shen:2009aa} in the presence of an updated cosmic ionizing background \citep{Haardt:2012aa}.

The recipes for star formation and SN feedback are the same in all the runs and are described in  \citet{Stinson:2006aa}. Gas particles must be dense -- namely have a density above the threshold $n_{\rm SF}$ (set to 100  atom~cm$^{-3}$  in E2k and 5 atom~cm$^{-3}$ in the other runs) -- and cool ({$T < T_{\rm max} = 1$--$3 \times 10^4$~K}) in order to form stars.  Particles which fulfill these requirements are stochastically selected to form stars according to $dM_*/dt = c^* M_{\rm gas}/t_{\rm dyn}$, where $M_*$ is the mass of stars created, $c^*$ is a constant star formation efficiency factor (set to 0.1 in all runs), $M_{\rm gas}$ is the mass of gas creating the star, and $t_{\rm dyn}$ is the gas dynamical time. Each star particle then represents a population of stars, covering the entire initial mass function (IMF; listed in Table~\ref{tab:tab0}).

Stars {more massive} than 8~M$_{\odot}$ explode as SNII. According to the ``blastwave feedback'' model of \citet{Stinson:2006aa}, feedback is purely thermal, as the blastwave shocks convert the kinetic energy of ejecta into thermal energy on scales smaller than those resolved by our  simulations. Once energy is ejected (the fraction of SN energy that couples to the interstellar medium is $\epsilon_{\rm SN} = 1.0$ in E2k and 0.8 in the remainder), particles receiving the energy are prevented from cooling for typically 10--50~Myr, with the cooling shut-off timescale being computed as the sum of the Sedov--Taylor \citep{Taylor:1950,Sedov:1959} and snow-plough phases in the ejecta \citep{McKee_Ostriker_1977}. By delaying the cooling{,} we model in a phenomenological way the unresolved effect of momentum and {energy input} by turbulent dissipation in the ejecta before they reach the radiative phase. The strength of feedback depends on the number of SNe produced, which {in turn is} governed by the IMF and, locally, by the star formation density threshold. 

The IMF in Eris, EBH, and Venus {is from} \citet{Kroupa:1993aa}, whereas {the IMF} in E2k {is the updated one from \cite{Kroupa:2001aa},} which yields about a factor of 2.8 more SNe {for the same} star formation rate. Furthermore, as explained in detail in \citet{Guedes:2011aa} and \citet{Mayer:2012}, the local star formation rate, and thus the local effect of SNe, can be boosted significantly by raising the star formation density threshold as the interstellar medium is allowed to become more inhomogeneous, an effect that saturates only at very high resolution and density thresholds, well above those resolved with cosmological simulations \citep{Hopkins_et_al_2012}. This implies that in E2k heating by SN feedback is boosted both globally and locally. We recall that E2k is a run that follows an extensive study of sub-grid parameters by running many different simulations with the same Eris-type initial conditions in order to determine the combination of parameters that yields realistic stellar masses in accordance with abundance matching at both high and low  redshift, these being shown in Table~\ref{tab:tab0}. Indeed in the original Eris suite the conversion of gas into stars was too efficient at high redshift, although final stellar masses at $z = 0$ are in agreement with abundance matching \citep[see also][]{Agertz:2014,Sokolowska:2016}. E2k also has  a richer inventory of physical processes, not only metal-line cooling but also a sub-grid turbulent diffusion prescription for both metals and thermal energy which allows mixing to be captured in SPH \citep{Shen:2009aa}. {In all the runs, metals come from SNI and SNII \citep{Stinson:2006aa}.}

The EBH  run improves the physical model in the simulations in a different direction as it includes prescriptions for the formation, growth, and feedback of supermassive black holes, and assumes ``quasar mode'' thermal feedback with Bondi--Hoyle--Lyttleton accretion (\citealt{bondi52,bondi44,hoyle39}; for more details, see \citealt{bellovary10,Bonoli:2016}), while all the rest of the sub-grid modeling and the cooling is identical to that of Eris. {\cite{Bonoli:2016} found that, except at $z > 3$, when major mergers occur, gas accretion onto the central supermassive black hole is always well below the Eddington rate. As a result,
radiative feedback from gas accretion is negligible. Radiative feedback affects only the very central region (within 1~kpc), resulting in the suppression of the bulge growth relative to Eris. The small bulge is likely the reason why, at $z<1$, the disk of EBH is more prone to instabilities, which cause the growth of a strong bar at $z<0.3$ \citep[more details can be found in ][]{Spinoso:2016}.}

The Venus simulation employs different initial conditions. The ``zoom-in''  was initialized using the {\sc music} code \citep{MUSIC}, rather than with {\sc grafic2} \citep[][]{Bertschinger_2001} as in the other cases, which allows a computationally more efficient topological identification of the Lagrangian subvolume for the refinement. {Eris and Venus both form at the intersection of four dark matter filaments, albeit their convergence pattern is different. In general, Venus experiences twice as many major mergers as the other runs, with its last major merger (defined as a merger with mass ratio $>0.1$ between the two galaxies) occurring
at $z = 0.9$, as opposed to $z = 3.1$ in the other
runs. While in Eris a central dominant halo assembles very early, in Venus multiple progenitors of comparable mass evolve separately for a long time, with one single halo and its associated galaxy only appearing after the last major merger at $z<0.9$. The amount of substructure at $z=0$ is also more abundant in Venus relative to Eris, both in the stellar and in the dark matter
component. In particular, a large satellite orbits around the primary galaxy in Venus even at late times, causing a perturbation on the main disk at pericenter passages, the last of which induces perturbations in the structure of the main disk as late as $z=0.24$.}

\section{Disk--bulge decompositions}\label{sec:bulge-disk-decomposition}

We begin {by} characterizing the structure of our simulated galaxies, {decomposing them into disk and bulge components by two} complementary methods. We perform the decompositions of the simulated spiral galaxies at the final redshift $z_{\rm end}$ of each simulation ($z_{\rm end} = 0$ for Eris, EBH, and Venus; $z_{\rm end} = 0.3$ for E2k){, considering} the stellar population in the spherical region {within 15 comoving kpc of} the minimum of the potential well of the galactic halo. We chose this radius as the limit, upon a visual inspection of the extent of the galaxies at $z = z_{\rm end}$ in the {stellar} density maps. {We do not identify or subtract particles that might belong to a stellar halo. However, we verified that at $z_{\rm end}$ our results are only mildly sensitive to a range of scale-height tresholds. Moreover, even when we
consider all stars within the spherical region, most
of the stellar halo is not included in our sample since
it would extend to larger radii \citep[see][]{Rashkov:2013}. Henceforth we use the term bulge to refer to
all stellar particles that do not belong to the disk.}

\begin{figure}
\vspace{-10.0pt}
\centering
\includegraphics[scale=.23]{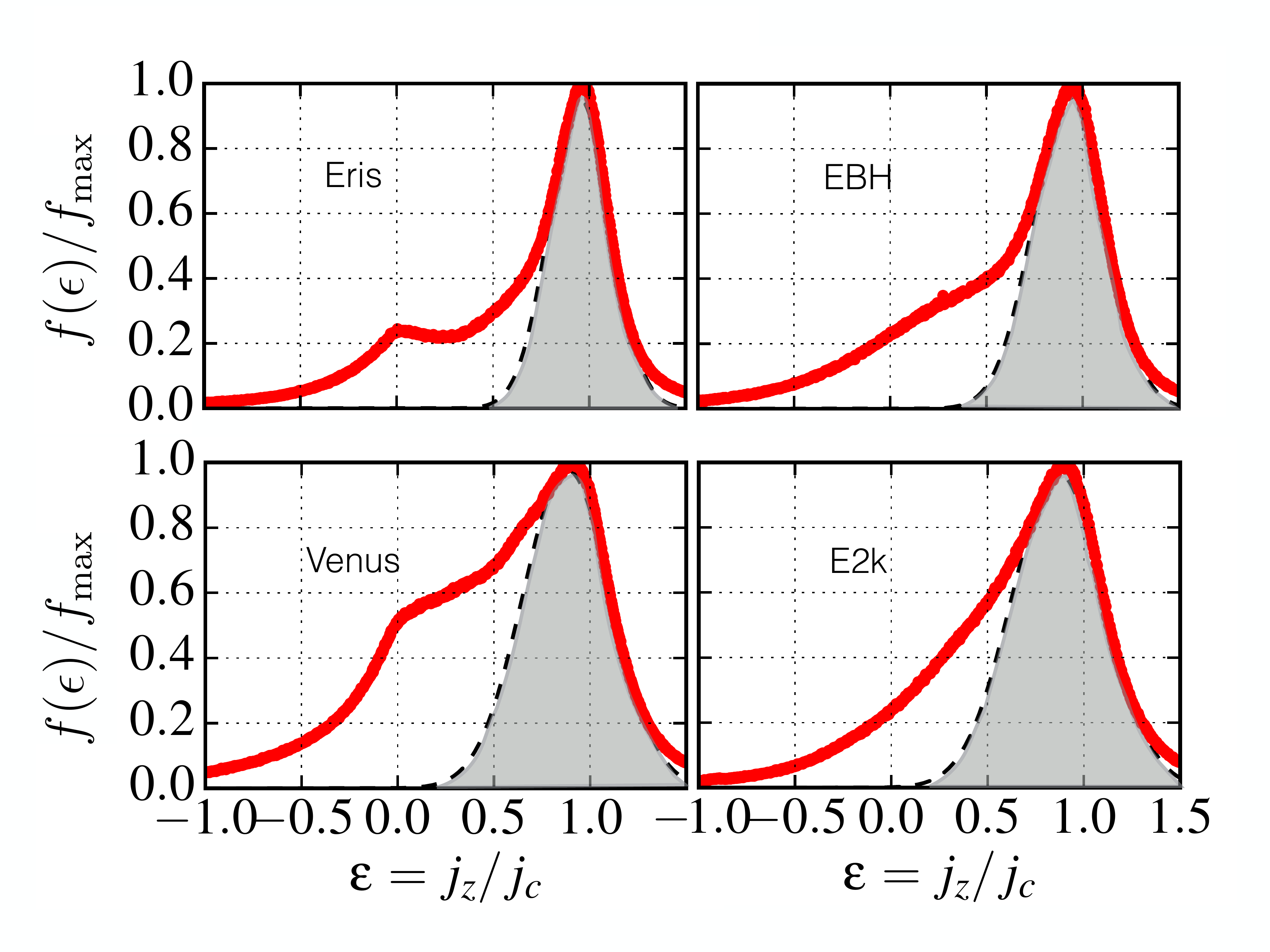}
\vspace{10.0pt}
\caption{The distribution of the circularity parameter $\epsilon$ for stellar particles in a galaxy at the most recent redshift (Eris, EBH, and Venus: $z_{\rm end} = 0$; E2k: $z_{\rm end} = 0.3$) is shown with the red line. The black dashed line denotes the Gaussian function obtained as a fit to the distribution right of and about the highest peak in the distribution ($\epsilon\simeq1$). All particles in the grey-shaded area are assigned to the disk.}
\label{fig:kin0}
\end{figure}

{The first decomposition method is based on kinematics.} In Figure~\ref{fig:kin0}, we show four distributions of the circularity parameter $\epsilon \equiv j_{\rm z}/j_{\rm c}$, where $j_{\rm z}$ is the $z$-component of the angular momentum vector of a stellar particle when the galactic disk lies in the $x$--$y$ plane, and $j_{\rm c}$ is the angular momentum of {a hypothetical} particle {at the same location on a circular orbit}.  One expects the distribution of the circularity parameter, {$f(\epsilon)\equiv \Delta N/\Delta \epsilon$ (where $\Delta N$ is the number of particles in a circularity bin $\Delta \epsilon$)}, to show two peaks in a typical spiral galaxy{:} one at $\epsilon \simeq 1$ {for} the disk, and one at $\epsilon \simeq 0$ {for the bulge}.

Using the information about the direction of the total angular momentum of all particles in our sample, we rotate the galaxies to appear face-on in the $x$--$y$ plane.  Then we compute the circularity parameters $\epsilon$ of all particles and plot their mass-weighted histogram, which is then normalized to the maximum value of the distribution, $f_{\rm max}$, as shown in  Figure~\ref{fig:kin0} (note that $f_{\rm max}$ represents the most probable value of the circularity).

The total distribution (red line) is the sum of the particles {in the disk and bulge components. Some of the galaxies in our sample exhibit non-axisymmetric features such as bars \citep[see the discussions in][]{Bonoli:2016, Guedes:2013}; therefore,} a standard picture of a \citet{Sersic_1963,Sersic_1968} {classical} bulge and a thin disk (or, equivalently, a sum of two Gaussian distributions of particles peaking at $\epsilon = 0$ and $\epsilon = 1$) is not applicable. However, as seen in Figure~\ref{fig:kin0}, all spiral galaxies in our sample have a distinct, close-to-Gaussian distribution for $\epsilon \gtrsim 0.7$. We thus fit a Gaussian function to each of these distributions for all particles with $\epsilon > 0.8$ at $z_{\rm end}$ to determine the mass and the angular momentum of the disk component, treating the rest as the {bulge}. The fits are shown with black dashed lines in Figure~\ref{fig:kin0}.

To determine the {bulge}-to-disk ($B/D$) and {bulge}-to-total ($B/T$) ratios, we first denote the Gaussian fit to the disk as $g(\epsilon)$ and define the weighting function $w \equiv [f(\epsilon)-g(\epsilon)]/f(\epsilon)$, which {is} the fraction of stellar particles in each bin of {the circularity} histogram {assigned} to the {bulge}. The $B/D$ ratios can be determined from

\begin{eqnarray}
D = \int_{\epsilon_{\rm min}}^{\epsilon_{\rm max}}[1-w(\epsilon)]M(\epsilon)d\epsilon, \label{eq:D} \\ 
B = \int_{\epsilon_{\rm min}}^{\epsilon_{\rm max}}w(\epsilon)M(\epsilon)d\epsilon, \label{eq:S}
\end{eqnarray}

\noindent {where $M(\epsilon)d\epsilon$ is the mass of all stellar particles with circularity between $\epsilon$ and $\epsilon + d\epsilon$. We set} $\epsilon_{\rm min}=-1.5$ and $\epsilon_{\rm max}=1.5$. The final values of ratios {$B/D$ and $B/T$ (}where $B + D = T$) for the $z=z_{\rm end}$ galaxies determined with this method are $(B/D)_{\rm kin} =(0.75, 0.51, 0.96, 0.69)$ and $(B/T)_{\rm kin}=(0.43, 0.34, 0.49, 0.41)$ for Eris, E2k, Venus, and EBH, respectively.

Although our primary method of decomposition in this paper is kinematic, we also decompose galaxies into {disks and bulges} photometrically {when} observed face-on (i.e., parallel to the galactic angular momentum vector). Comparing the results of these two methods may prove useful for determining the uncertainties on the properties derived from the 2D quantities (surface brightness/surface density) vs. those based on the 3D kinematic information. 

The mock data, i.e. the surface density profiles for the stars, are calculated {as before,} for a sphere of $r = 15$~comoving kpc around the centers of galaxies. The fitting function is a combination of the surface density of an exponential disk, $\Sigma_{\rm d}$, and a S{\'e}rsic {bulge}, $\Sigma_{\rm b}$:

\begin{equation}
\begin{multlined}
\Sigma_*(r) = \Sigma_{\rm d}(r)+\Sigma_{\rm b}(r) = 
\Sigma_{\rm d,0} \exp{\left( -\frac{r}{R_{\rm d}}\right)} +  \\ \Sigma_{\rm b,0}\exp{\left[ -b_{\rm n} \left(\frac{r}{R_{\rm b}}\right)^{1/n}-1\right]},
\end{multlined}
\label{eq:sigma}
\end{equation}

\noindent where $n$ is the S{\'e}rsic index, $R_{\rm d}$ and $R_{\rm b}$ are the scale radii of the two profiles, $\Sigma_{\rm d,0}$ and $\Sigma_{\rm b,0}$ are the {central surface densities}, and $b_{\rm n} = 1.9992\,n - 0.3271$ \citep{Capaccioli:1989}. Initially, we determine parameters with a least-squares method, whose best-fit parameters serve as initial guesses to the more sophisticated Metropolis-Hastings Markov chain Monte Carlo method \citep{Metropolis:1953,Hastings:1970} with $10^6$ realizations. We {have modifed this algorithm to} accept only those sets of parameters  that result in the best fit to our data. 

\begin{figure}
\vspace{-10.0pt}
\centering
\includegraphics[scale=.8]{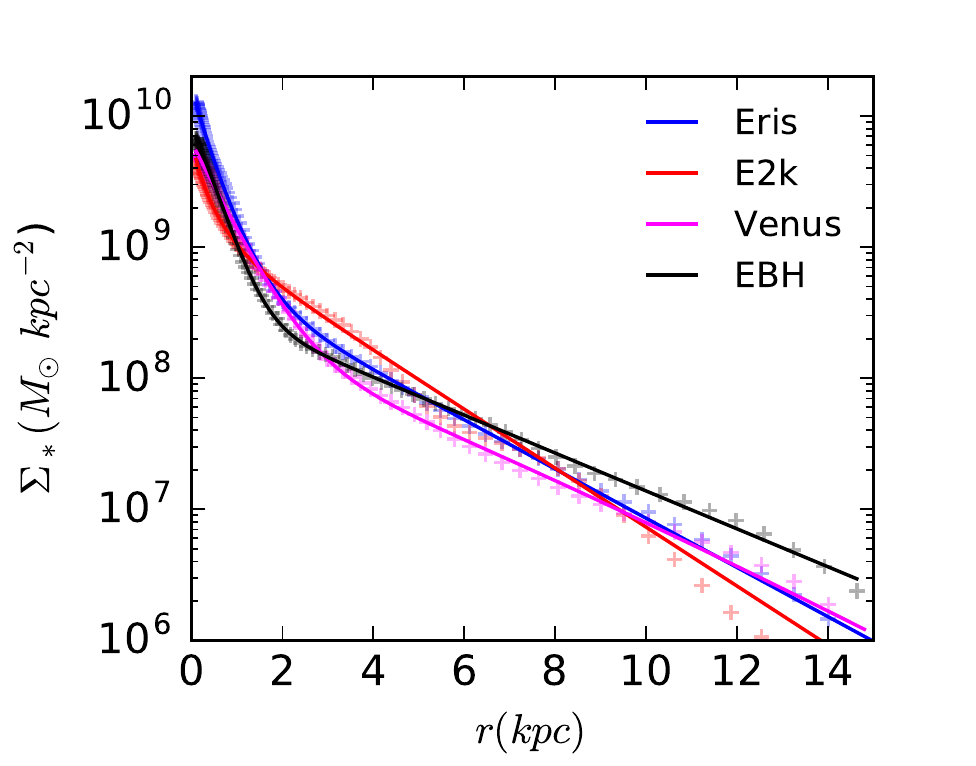}
\vspace{10.0pt}
\caption{Best-fit functions to the surface-density profiles of our sample of simulated galaxies. The actual surface-density profiles are denoted with crosses, whereas the solid lines show the result of the fitting functions (see Equation~\ref{eq:sigma}). The calculations are performed at $z = z_{\rm end}$.} 
\label{fig:bestfit}
\end{figure}

The results are shown in Figure~\ref{fig:bestfit}, with the best-fit parameters and estimates of bulge-to-disk ratios {listed} in Table~\ref{tab:tab1}. We computed the bulge-to-disk and bulge-to-total ratios according to $B=\int_0^{\infty} \Sigma_{\rm b}(r) 2\pi r dr$ and $D=\int_0^{\infty} \Sigma_{\rm d}(r) 2\pi r dr$. When compared to the kinematic estimates, shown again in Table~\ref{tab:tab1}, these ratios are in near-perfect agreement for Eris and Venus, {but are different for EBH and E2k}. Nevertheless, the relative values of $(B/T)_{\rm ph}$ and $(B/T)_{\rm kin}$ ratios are consistent between these two methods, i.e. the sequence of runs with increasingly prominent bulge is in both cases: E2k, EBH, Eris, and Venus. Quantitatively, the differences between the two methods are in the 10--30\% range for all galaxies except E2k, for which the difference is about a factor of 3. 

{Each decomposition method has its own limitations, which can affect derivations of the disk-to-bulge ratio and thus measurements of properties associated to it \citep[][]{Abadi:2003, Scannapieco:2010, Guidi:2015}. Here,} the differences arising in the bulge-to-disk ratios between the photometric and kinematic decomposition methods are strongest for the galaxies with the most prominent bars, which illustrates how the accuracy of the photometric method {depends} on the type of {bulge. Neglecting to separate a bar may lead to an overestimate of the bulge luminosity by 50\% \citep{Gadotti:2008}.} The kinematic method enables {us} to distinguish between the material of high and low velocity dispersion, while the S\'{e}rsic index measures only the curvature of the surface density profile, and does not necessarily describe an object that is truly round or flat in 3D. We note that galaxies {with} bars might turn out to be morphologically closer to pure disk objects (E2k), $Sc/Sd$ galaxies (EBH), or $Sb/Sc$ galaxies (Eris), as we do not {distinguish bars from bulges} in this paper.

The more pronounced difference in the case of E2k, in which the photometrically {defined} bulge is much less prominent than {the kinematically defined one}, highlights a peculiarity of the galaxy with stronger feedback. By low redshift, this galaxy has acquired a stellar component with no clear separation between disk and {bulge} (see Figure~\ref{fig:kin0}). However, when the surface brightness profile is inspected, it is almost a single exponential up to less than 1~kpc from {the center}, which is typical of very {late-type galaxies}. The central steepening of the profile inside 1~kpc is highly correlated with the growth of a bar around $z = 1$ and below, as we checked that the inner profile is flatter earlier on. We argue that E2k is essentially a bulgeless\footnote{{Here we define a bulgeless galaxy as one with no detectable extended central component separate from the disk. According to this terminology, a galaxy with a nuclear star cluster such as M33 would
be classified as bulgeless.}}, barred disk galaxy, or equivalently that the bar makes up for a large fraction of what we identify as the bulge with the photometric method. The tendency of stronger SN feedback to suppress bulge formation \citep[e.g.][]{Hopkins:2014,Keller:2014} is expected since ejective feedback removes low-angular momentum baryons by means of outflows \citep[see][]{Governato:2010,Brooks:2013}. The tendency of this galaxy to have a kinematically hotter stellar component is also likely an effect of feedback on the galactic structure, which will be studied in Paper~II, but may also reflect the presence of a rather prominent bar which is expected to induce non-circular motions in the stellar component.

We note that, in observations, the quoted $B/T$ ratios are normally obtained by applying the fits and decomposition to the surface brightness profile in a given {photometric} band, rather than to the surface {mass} density. Depending on the band, the difference in the relative weight of the bulge {and disk} can be small {or large}, with variations of up to a factor of 3 between the B and the I band, depending on the {stellar age distributions} \citep{Graham_Worley_2008}. In general, the bulge, which is composed of an older stellar population than that of the disk, will be fainter in optical bands relative to the {disk.} Hence our estimates of the relative contribution of the two components, which are based on actual mass density, should be considered as an upper limit. Indeed in the case of Eris, the $B/D$ {mass} ratio we quote here is higher by a factor of 2 with respect to the I-band $B/D$ found with {\sc galfit} \citep{Peng:2002,Peng:2010} after post-processing with the {\sc sunrise} radiative-transfer code \citep{Jonsson:2006,Jonsson:2010}, including dust reddening \citep{Guedes:2011aa}. This supports the notion that, photometrically, E2k is an almost bulgeless galaxy as its $B/D$ would be $< 0.1$ in optical bands.

\tabletypesize{\scriptsize}
\begin{deluxetable}{c|c|c|c|c}
\tablecaption{Bulge--disk decomposition parameters}
\tablenum{2}
\tablehead{\colhead{Parameter} & \colhead{Eris} & \colhead{E2k} & \colhead{Venus} & \colhead{EBH}}
\startdata
$\Sigma_{\rm d,0}$		& $6.43 \times 10^8$	& $1.29 \times 10^9$	& $6.12 \times 10^9$	& $3.85 \times 10^8$ \T \B \\
$R_{\rm d}$		& 2.31				& 1.93				& 0.62				& 2.44 \B \\
$\Sigma_{\rm b,0}$	& $2.43 \times 10^9$	& $5.67 \times 10^8$	& $4.97 \times 10^7$	& $1.80 \times 10^9$ \B \\
$R_{\rm b}$		& 0.70				& 0.70				& 4.84				& 0.67 \B \\
$n$				& 1.14				& 1.35				& 0.85				& 0.88 \B \\
\hline
$(B/T)_{\rm ph}$				& 0.41				& 0.11				& 0.47				& 0.29 \T \B \\
$(B/T)_{\rm kin}$				& 0.43				& 0.34				& 0.49				& 0.41 \T \B \\
\hline
$(B/D)_{\rm ph}$				& 0.70				& 0.13				& 0.89				& 0.41 \B \\
$(B/D)_{\rm kin}$				& 0.75				& 0.51				& 0.96				& 0.69 \B \\
\hline
$j_{\rm d}$ & 784.6 & 762.6  & 618.4& 952.0\\
$j_{\rm b}$ & 143.0 &  133.0 & 101.2 & 139.3\\
$j_{\rm star}$ & 511.0 & 547.4 & 363.3 & 613.2 \\
$j_{\rm gas}$ & 1620.2 & 916.4 & 1541.4& 1829.6\\
\enddata
\tablecomments{Best-fit values of the parameters of the photometric decomposition (see Equation~\ref{eq:sigma}). For a comparison, we also show the results of the kinematic decomposition -- $(B/T)_{\rm ph}$ vs. $(B/T)_{\rm kin}$ and $(B/D)_{\rm ph}$ vs. $(B/D)_{\rm kin}$. We add the values of the specific angular momenta calculated at $z_{\rm end}$ for: $j_{\rm d}$ -- stars in the disk; $j_{\rm b}$ -- stars in the bulge; $j_{\rm star}$ -- stars in the disk and bulge; $j_{\rm gas}$ -- cold ($T<10^4$~K) gas in the galaxy. The units of $\Sigma$, $R$, and $j$ are M$_{\odot}$~kpc$^{-2}$, kpc, and km~s$^{-1}$~kpc, respectively.}
\label{tab:tab1}
\end{deluxetable}

{ We provide a more complete discussion of bulge classication in the Appendix, where we investigate,
for each of our simulated galaxies, whether it harbours a pseudobulge or a classical bulge, based on six independent observationally and physically motivated criteria. Some of
the methods yield contradictory results, which underlines
the importance of examining as many criteria as
possible. In essence, we conclude that, at $z_{\rm end}$, our sample includes composite, classical, pseudo-, and peanut bulges (in Eris, Venus, E2k, and EBH, respectively).}

\section{$j_*$--$M_*$ diagrams}\label{sec:j-M-diagrams}

In this section, we use the outcome of the preceding analysis to determine the specific angular momentum of the disk and bulge, as well as that of the overall stellar and gas components. In this way, we can compare the scaling relation between stellar mass and specific angular momentum with those of observed galaxies, as well as study the evolution of such relations from high to low redshifts. This is particularly relevant for the interpretation of the $j_*$--$M_*$ diagram{,} which has been proposed as an alternative to the Hubble sequence {\citep{Fall:1983,Romanowsky:2012}}.

We calculate the specific angular momentum of our sample of galaxies at $z = z_{\rm end}$ for the entire galaxies and for their separate components. The specific angular momentum vector of particle species $k$ is defined as

\begin{equation}
{j}_k = \frac{\sum_{i} m_{k,i} {r}_{k,i} \times {v}_{k,i}}{\sum_i m_{k,i}},
\end{equation}

\noindent where the sums are over each particle $i$. The particles within each histogram bin (Figure~\ref{fig:kin0}) are assigned to a {bulge} or a disk randomly but in numbers determined by the weight function $w$ defined in Section~\ref{sec:bulge-disk-decomposition}. 

{The random selection is applied in the interest of simplicity. Although one could additionally classify particles based on their distance from the galactic center or their stellar ages, these simple criteria also bear a degree of arbitrariness -- big bulges can extend far from the galactic center (e.g. the Sombrero galaxy), and disks may also include old stars \citep[see Figure~5 of ][]{Guedes:2013}. This random selection matters most in the transition region of the circularity distribution between the bulge and the disk, thereby it affects a small number of particles. We have verified that assigning the particles in the transition region differently to the bulge and disk (e.g. based on a variety of distance thresholds) does not change the relative weight of disk and bulge by more than 10\%.}

\begin{figure*}
\vspace{-10.0pt}
\centering 
\includegraphics[scale=0.9]{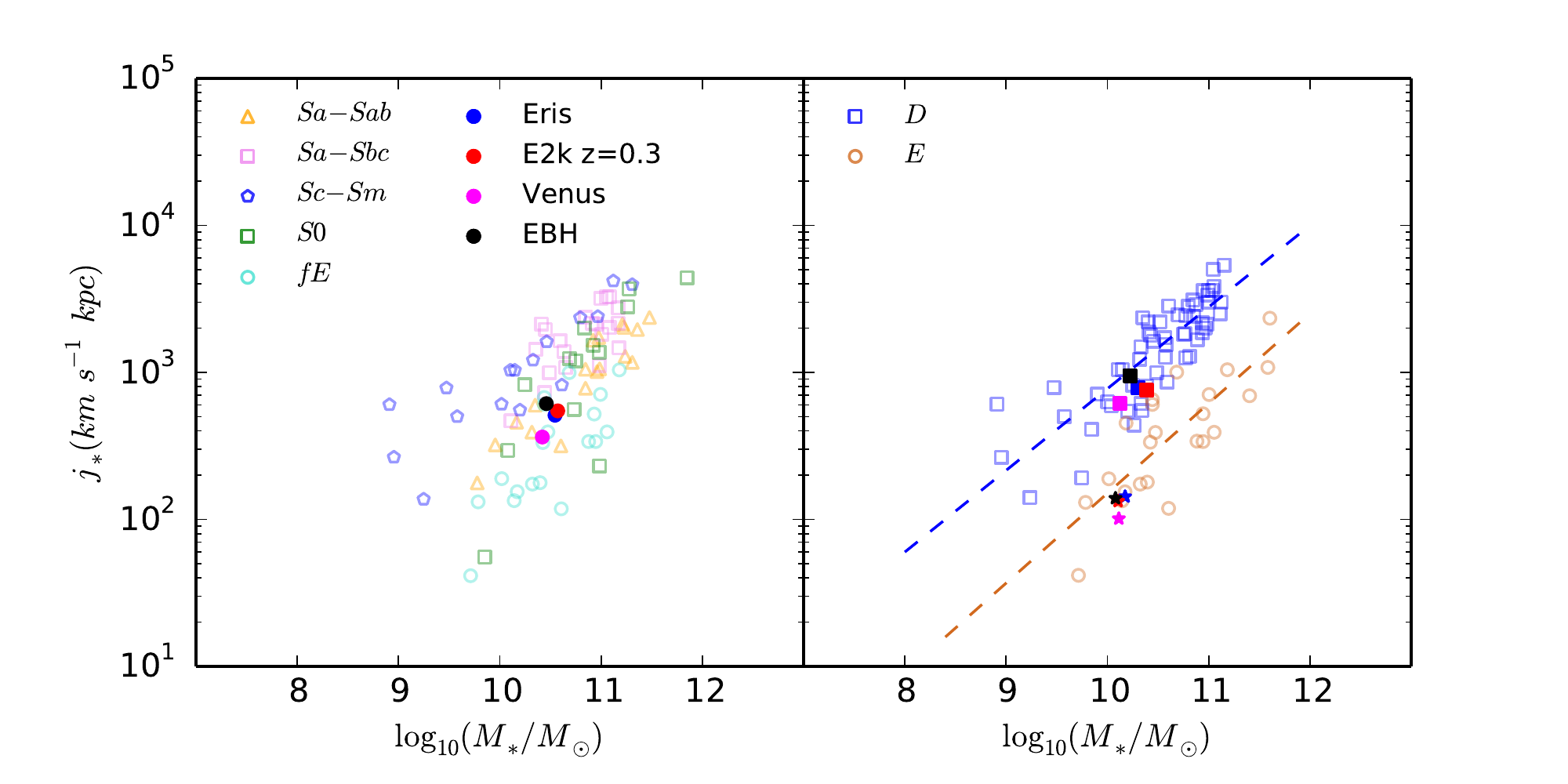}
\caption{Specific angular momentum--mass ($j_*$--$M_*$) diagrams for stars of the simulated galaxies vs. the sample of \citet{Fall:2013}. Left. The comparison of the total specific angular momentum of the simulated galaxies with the observed galaxies of various morphological types. Right. Simulated galaxies are kinematically decomposed into disks {(filled squares) and bulges (filled stars)} and then compared with the subsample of observed pure disk galaxies (D) and ellipticals (E).}
\label{fig:jstK}
\end{figure*}

Our results are set against two samples of observed galaxies. In the left panel of Figure~\ref{fig:jstK}, we compare the total specific angular momentum of the stars in our galaxies with a sample of galaxies classified {according to their morphologies \citep{Romanowsky:2012,Fall:2013}}. The location of our simulated galaxies  on that diagram is well-aligned with the population of observed disk-dominated galaxies. Furthermore, in the right panel, we compare the individual components, i.e. {bulge}s (marked as stars) and disks (marked as squares), with the sample of pure {disks and elliptical galaxies}. Upon the decomposition, it is evident that all simulated galaxies consist of a disk component with high specific angular momentum and a {bulge} component with low specific angular momentum. Moreover, in that respect, the disks of the simulated galaxies are in perfect agreement with what is expected of bulgeless galaxies, and their {bulge}s also agree well with what is found for ellipticals. The ratio of specific angular momentum between the corresponding components ranges from 5.5 to 6.8 (see also Table~\ref{tab:tab1}).

{Morphologically, bulges may be regarded as elliptical galaxies embedded in disks \citep{Kormendy:2004}}. In practice, this interpretation might change if a bulge was formed by secular processes (so-called pseudobulge, discussed more in {the Appendix}). Nevertheless, both the total $j_*$--$M_*$ diagrams, as well as the dichotomy in the distribution of the specific angular momentum of the components, confirm that the simulated galaxies do not suffer from the overcooling problem or the angular momentum catastrophe (see Section~\ref{sec:introduction}), and can be regarded as good laboratories for in-depth studies of the angular momentum evolution. 

\begin{figure*}
\vspace{-5.0pt}
\centering
\includegraphics[scale=0.23]{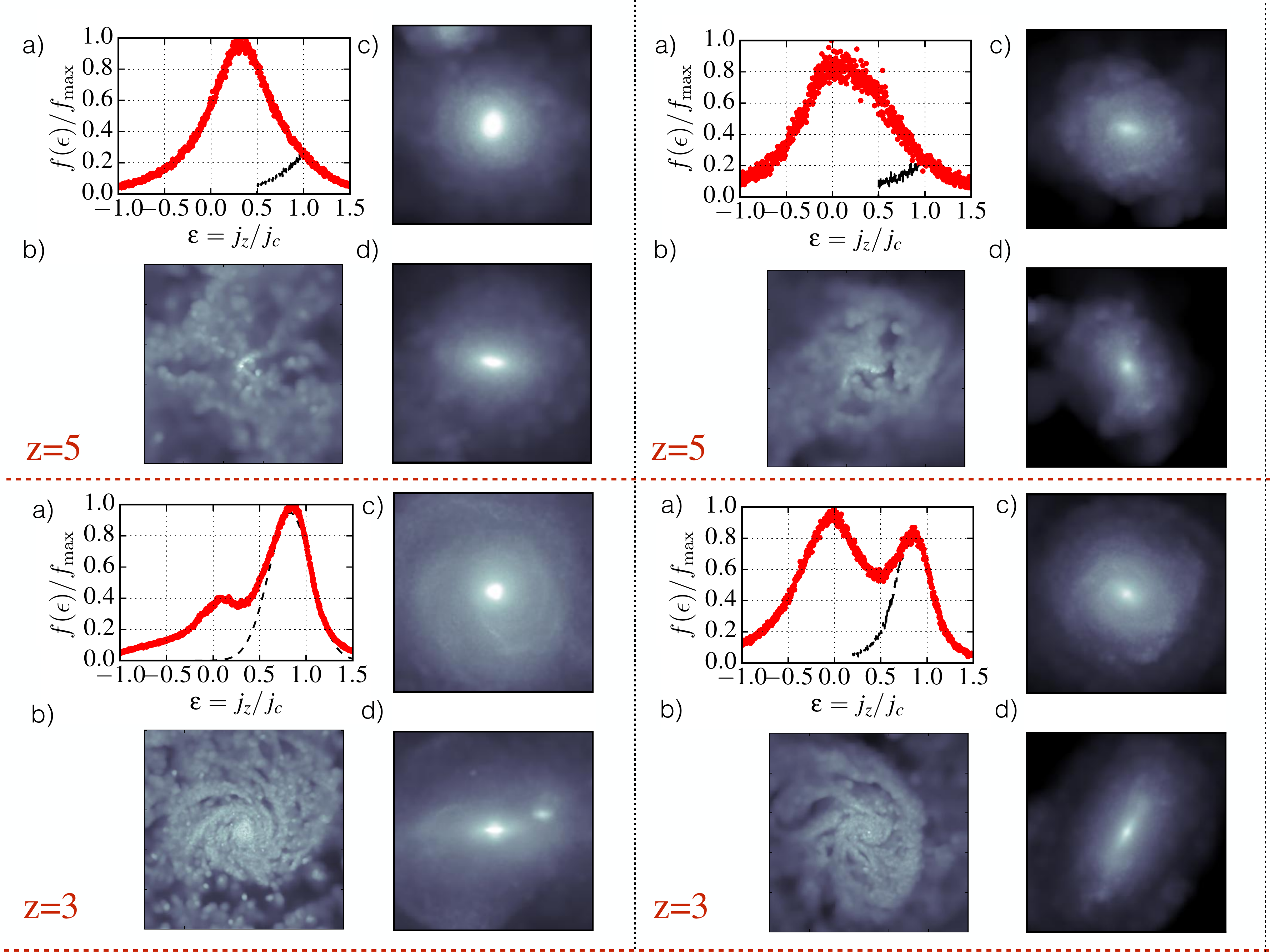}
\includegraphics[scale=0.23]{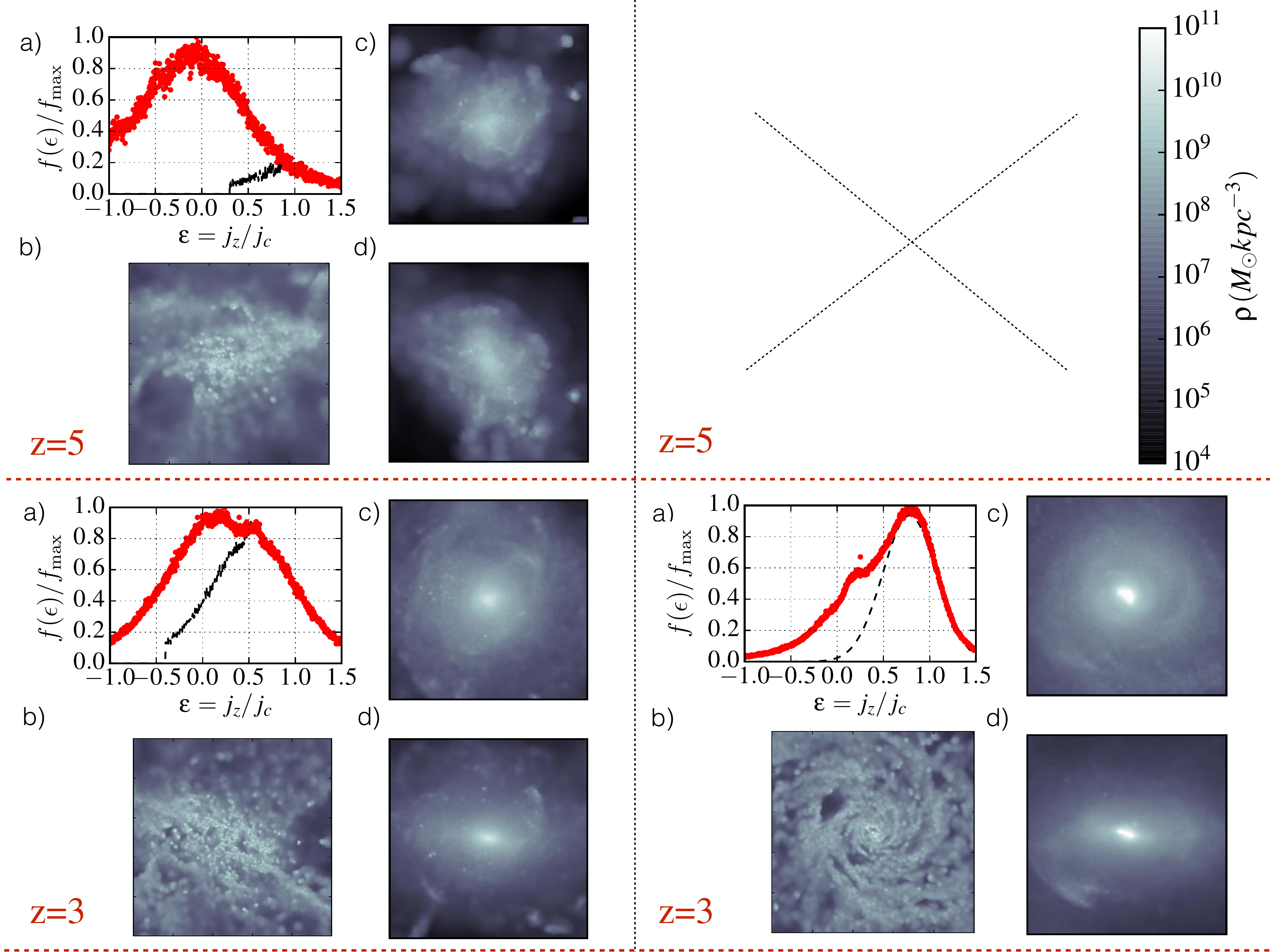}
\includegraphics[scale=0.23]{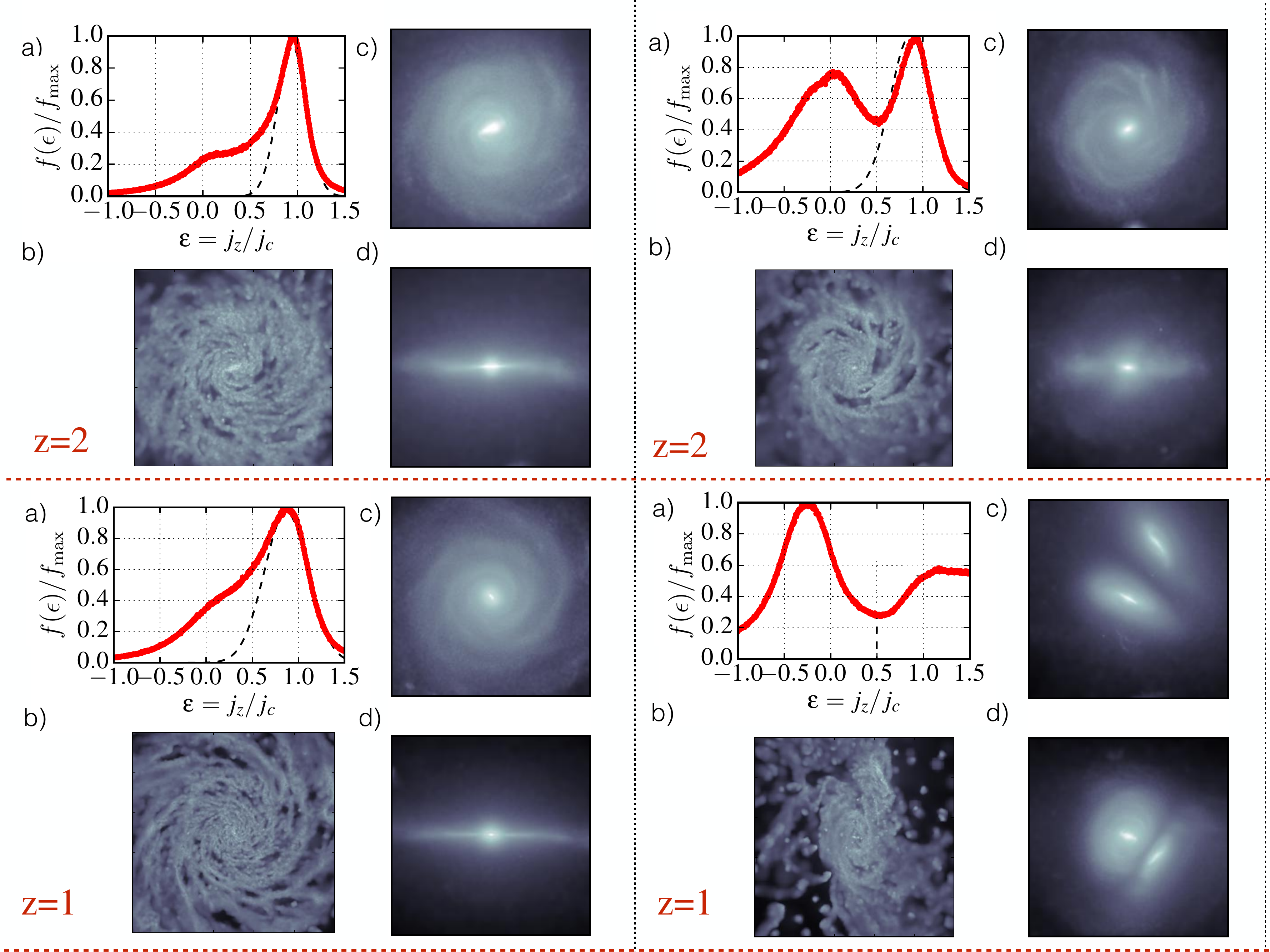}
\includegraphics[scale=0.23]{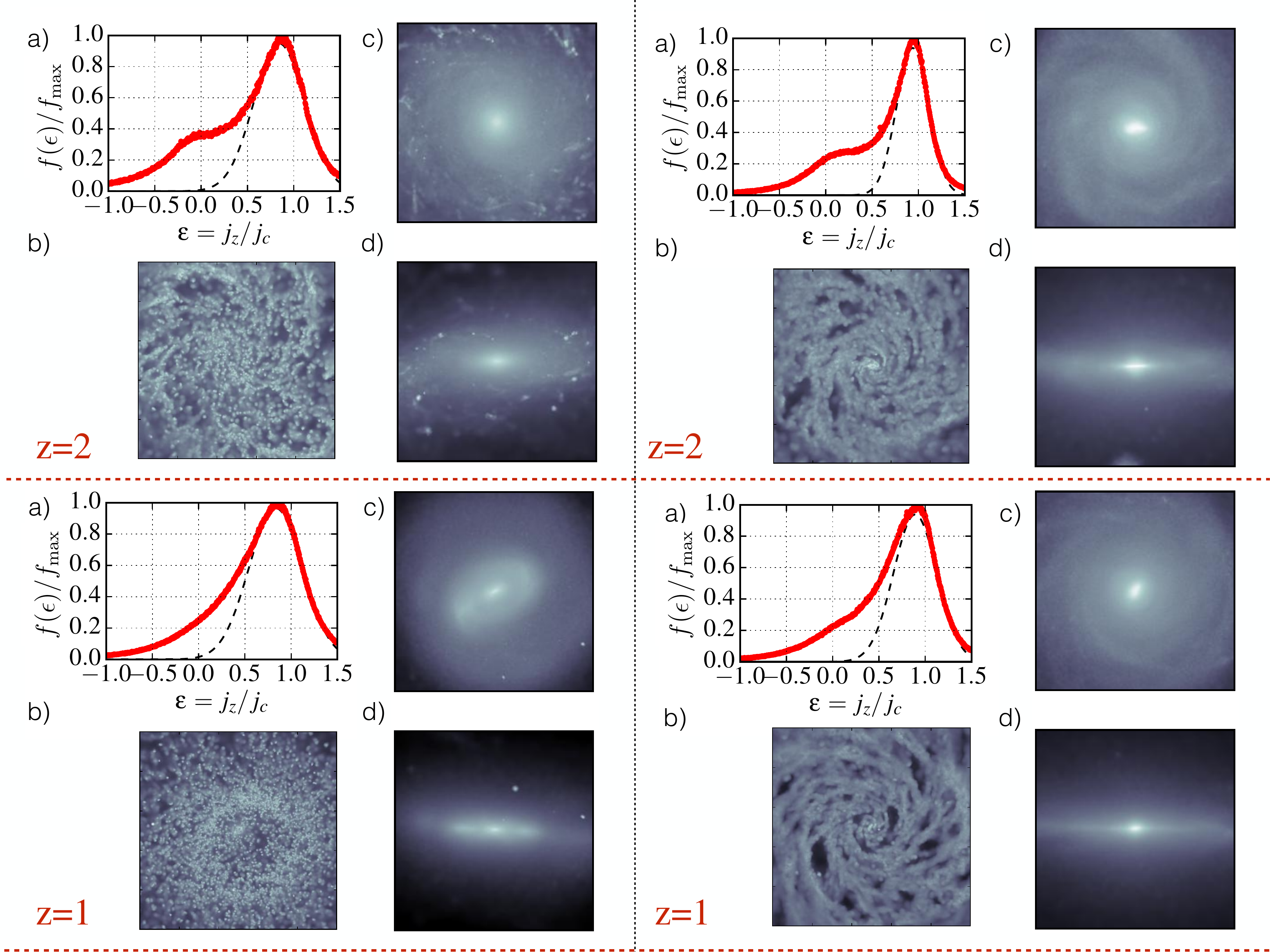}
\includegraphics[scale=0.23]{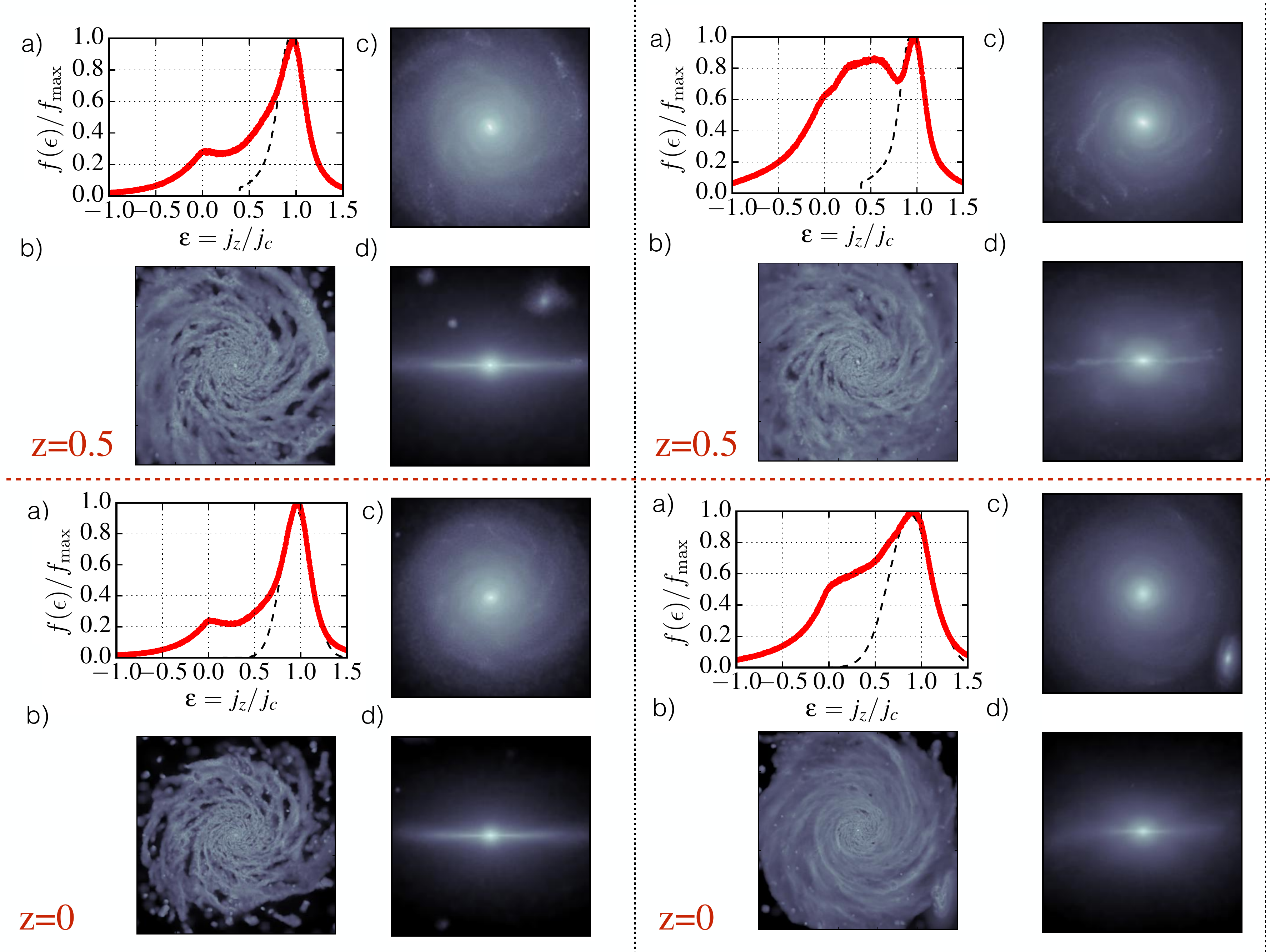}
\includegraphics[scale=0.23]{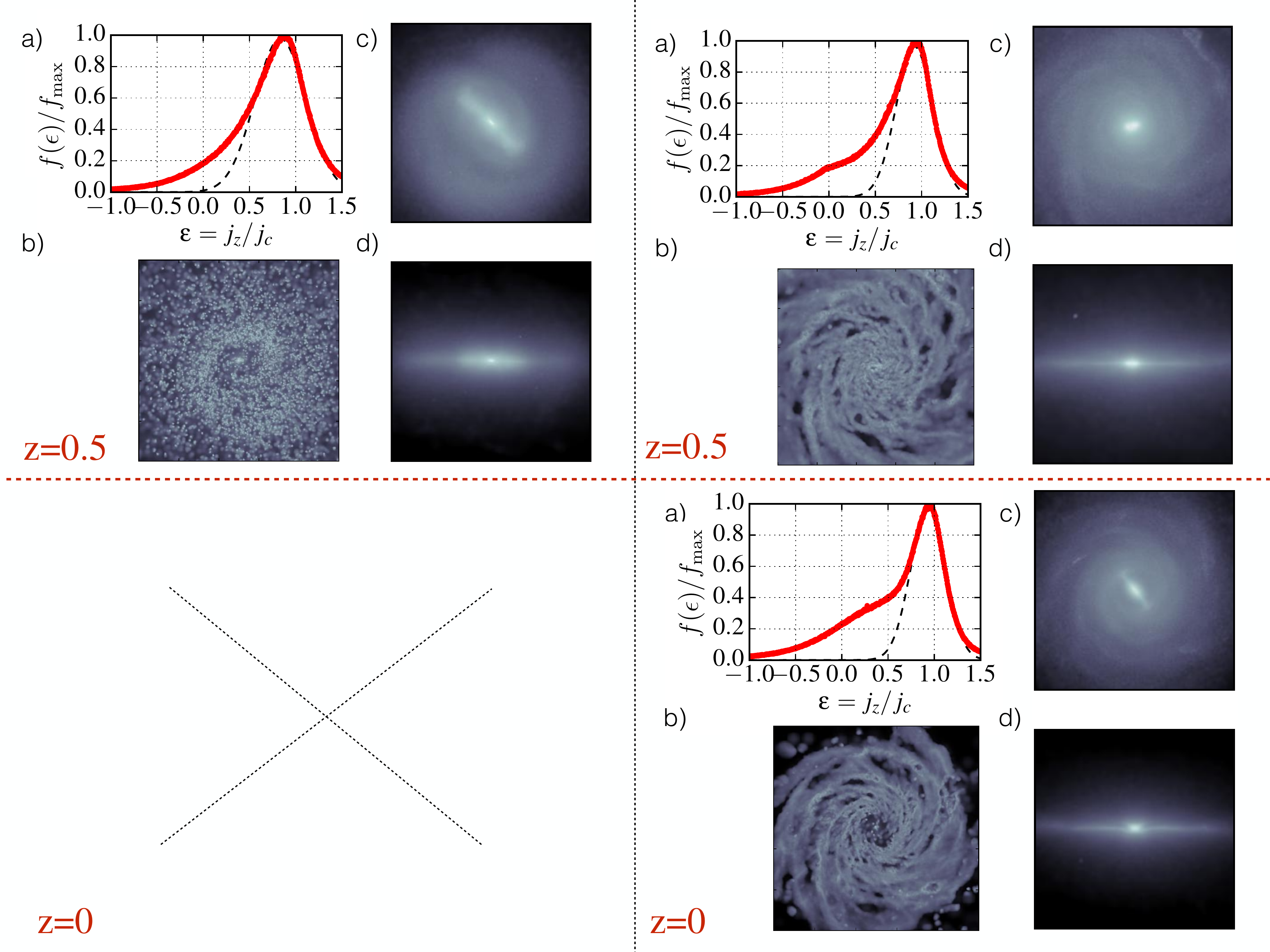}
\vspace{20.0pt}
\caption{A diagram linking the morphology of a galaxy at various stages of its lifetime with its kinematics. Each column corresponds to a different run -- going from left to right: Eris, Venus, E2k, and EBH. Every row shares the same redshift. Crosses are placed whenever an output of a run is missing. Each piece of a matrix contains the following information: a)~distribution of the circularity parameter in a galaxy; b)~gas density map of a galaxy oriented face-on; c)~stellar density map of a galaxy oriented face-on; d)~stellar density map of a galaxy oriented edge-on. Total circularity distributions are colored in red, whereas their sub-distributions assigned to the disks are marked in black. Every image has a width of 30~comoving kpc.} 
\label{fig:diagrams}
\end{figure*}

In what follows, we investigate the evolutionary tracks of the galaxies and their components on the $j_*$--$M_*$ diagram, {which has been proposed as a physically motivated classification scheme alternative to the Hubble sequence}, and also address their dependence on the $B/T$ ratio. To do so, first we need to decompose our sample of galaxies at various redshifts. {This is a non-trivial task at high redshift}, given the complexity of {galactic structure}, tidal interactions, and frequent mergers. We thus study kinematic diagrams along with the morphology of both gas and stars in order to properly interpret the data. Our results are presented in Figure~\ref{fig:diagrams}. 

In general, upon combining the circularity diagrams with the morphological data-set, in most of the cases the $B/T$ decomposition is straightforward. The same procedure as in the case of the $z = z_{\rm end}$ galaxies is applied, i.e. based on finding the thin disk in a sphere of 15~comoving kpc encompassing the galaxy. Whenever the fitting of a Gaussian fails (e.g. for both Venus and E2k at $z = 3$), we identify the peak of the circularity distribution that should correspond to the disk, i.e. near $\epsilon = 1$, and then characterize the disk as the ensemble of stars distributed symmetrically around the circularity peak (see for example view $a$ for E2k at $z = 3$ in Figure~\ref{fig:diagrams}).

All galaxies exhibit a similar morphology at $z = 5$, i.e. they appear to be ellipsoids rather than flat extended disks, although in the inner 1--2~kpc a flat disk-like component is already discernible. This and the fact that their {circularity distributions peak close to $\epsilon = 0$, lead to the classification of these galaxies as bulge}-dominated. The disky component is approximated by mirroring the distribution around {$\epsilon = 1$} as explained above, but with the peculiarity that at this stage there is no peak at high circularity yet. 

We caution that the galaxy structure at this redshift might suffer from resolution limitations, as the disk scale length would correspond to only a few gravitational softenings at this epoch (disk sizes are expected to be about an order of magnitude smaller simply from the scaling of the halo virial radius with redshift in a $\Lambda$--CDM cosmology, see e.g. \citealt{Mo:1998}). Indeed, recent simulations with much higher resolution (tens of pc)  that stop at $z > 5$ do find a prominent rotating disk in halos of masses only a few times larger than ours already at $z = 8$ \citep{Fiacconi:2016, Pallottini:2016}. However, these simulations also find that the disk is thick and turbulent, resulting in $v/\sigma < 2$ for the most part (where $v$ stands for the magnitude of the velocity vector and $\sigma$ is the total velocity dispersion), which supports the notion that the galaxy would be classified as {bulge}-dominated based on our criteria. An early phase in which a turbulent gas disk results in a thick primitive stellar disk was already pointed out in \citet{Bird:2013aa}.

In Figure~\ref{fig:diagrams}, the circularity diagrams {for} Eris and EBH have a dominant rotating disk around $\epsilon \simeq 0.8$, and a secondary peak near $\epsilon = 0$--0.1 at $z = 3$. A gaseous disk is evident{,} and the edge-on view  of stars appears flattened, although we witness signs of tidal disturbances from frequent interactions. This suggests that the galactic structure is {continually} evolving at this epoch and hard to characterize in a simple way. Most of the mass of Venus and E2k has a low circularity parameter peaking at {$\epsilon \sim 0$}, whereas the secondary peak is lower. The gaseous disk of Venus {begins to exhibit a global rotation pattern, in contrast to E2k, which probably reflects} the stronger effect of feedback on gas dynamics.

By $z = 2$, all galaxies are already dominated by a thick disk. The face-on stellar density maps reveal spiral structures present in all of them. By this time, Eris, E2k, and EBH have already entered a quiescent phase past the last major merger. Venus, on the other hand, experiences another major merger at $z \sim 1$. In this case, as there are two interacting galaxies at very small separation, the decomposition of the system is somewhat arbitrary. The circularity diagrams show two clear peaks, one at about {$\epsilon = 1$} and the other at {$\epsilon \simeq -0.2$}, thus we choose to cut the distribution at the minimum between the two peaks, i.e. at around $\epsilon = 0.5$. Despite the ongoing stellar merger (views $c$--$d$), the gaseous disk appears flat ($b$). The spiral structure of E2k vanished{,} giving way to a prominent bar. The circularity distribution, although strongly asymmetrical, shows only one peak near {$\epsilon = 0.8$}. 

From $z = 0.5$ to 0, the galaxies generally appear similar. The triple-component distribution of Venus settles into {one with two components} by $z = 0$ with an extended thin disk (in both the gaseous and stellar matter) and a massive stellar bulge. EBH develops a bar {\citep{Spinoso:2016} which appears less} prominent than the one in E2k in terms of size relative to the disk itself.

\begin{figure*}
\vspace{-10.0pt}
\centering
\includegraphics[scale=.5]{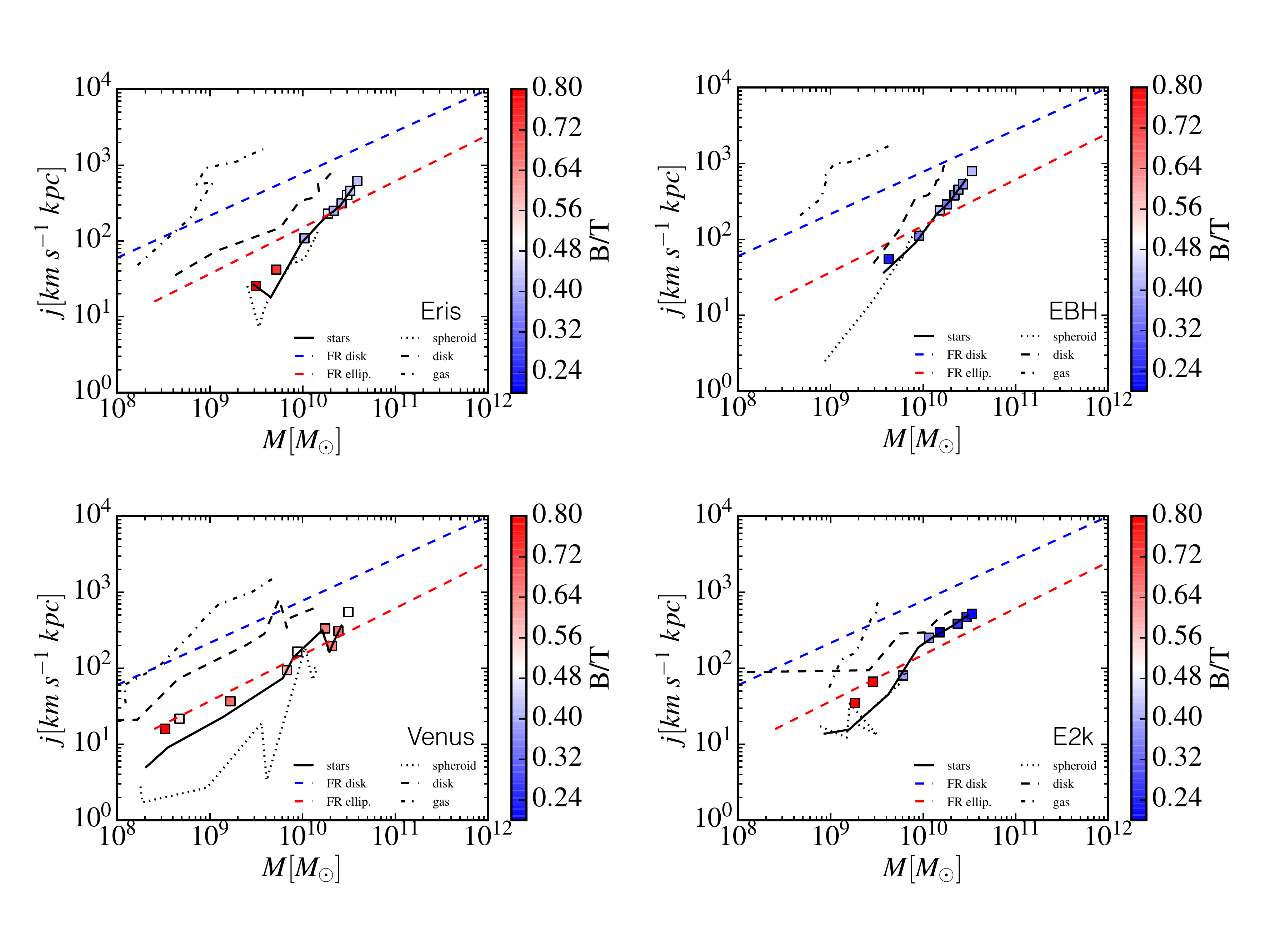}
\vspace{-30.0pt}
\caption{Specific angular momentum evolution of various components on the $j$--$M$ diagram. The gas considered here is cold (i.e. $T < 10^4$~K). ``FR disk'' and ``FR ellip.'' denote best-fit tracks for disk galaxies and ellipticals of \citet{Fall:2013}. Each data-point corresponds to the following redshifts (left to right): $z=(5,4,3,2,1.5,1,0.7,0.5,0)$ and represents the joint specific angular momentum of cold gas and stars (for EBH and E2k, respectively, the $z = 5$ and $z = 0$ data-points are missing). The color-coding ascribed to the data-points reflects the $B/T$ ratios of the galaxies at a given redshift.}
\label{fig:jtime}
\end{figure*}

\begin{figure*}
\vspace{-10.0pt}
\centering
\includegraphics[scale=.7]{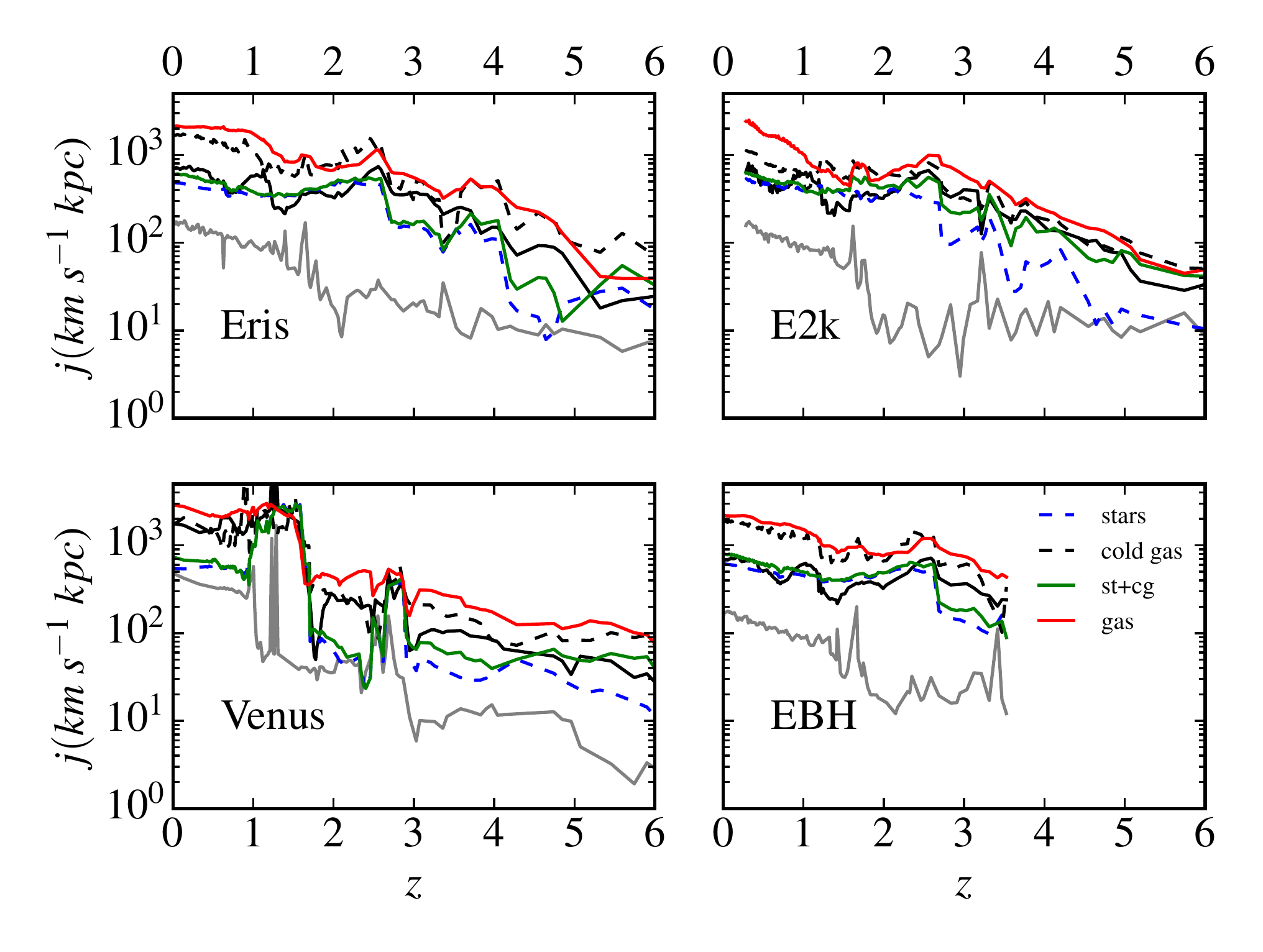}
\caption{Evolution of the specific angular momentum of the different components of our four galaxies: stars, cold gas ($T < 10^4$~K), all gas, and stars with cold gas within the virial radius as a function of redshift. The solid, black and grey lines represent the specific angular momentum for the dark matter within the virial radius and 10\% of the virial radius, respectively. Available data for EBH and E2k exist only for $z < 4$ and $z > 0.3$, respectively.}
\label{fig:jdm}
\end{figure*}

With a clear picture of morphological fluctuations in our sample of galaxies, we can proceed to quantify the magnitude of the specific angular momentum vector at various time steps. Figure~\ref{fig:jtime} shows evolutionary tracks {in the $j$--$M$ diagram of our four simulations for their} cold gas mass ($T<10^4$~K, dash-dot line), total stellar mass (solid line),  stellar mass in the {bulge} (dotted line), stellar mass in the disk (dashed black line), and joint total stellar and total cold gas mass (squares) at nine redshifts {[$z=(5,4,3,2,1.5,1,0.7,0.5,0)$; for EBH and E2k, respectively, the $z = 5$ and $z = 0$ data-points are missing]}. The data-points are color-coded with the $B/T$ ratios. The two diagonal dashed lines represent the relationship from \citet{Fall:2013} of $j_*\propto M_*^{\alpha}$ with $\alpha \sim 2/3$ for disk galaxies (in blue) and ellipticals (in red).

The specific angular momentum of cold gas is substantially higher than that of the stellar component ({reaching} values of order $10^3$~km~s$^{-1}$~kpc){, consistent} with observations \citep{Obreschkow:2014}. At all times, disks have a higher specific angular momentum than {bulge}s. At high redshift, all stars have a very low specific angular momentum that evolves below the line of ellipticals. After $z=3$, they move over that line and gradually get closer to the line of disks. Eris and EBH evolve {on a track with $\alpha = 1.4$, substantially steeper than one with $\alpha = 2/3$}. E2k, however, evolves on a track parallel to the one {with $\alpha = 2/3$}. Venus is a more complex case: it initially evolves approximately {along the sequence of elliptical galaxies} and, after several fluctuations, its track steepens in the final stages, after the last major merger ($z < 0.7$).

{In general, Eris, EBH, and E2k become disk-dominated galaxies past $z=4$ and evolve into galaxies
of different morphological types. Despite the fact that
our sample is too small to allow us to generalize, we
note that the galaxy with a classical bulge at $z=0$
has a lower stellar specific angular momentum than those
pseudo/composite bulges in Figure~\ref{fig:jstK}, and this is also true for the evolutionary tracks of Figure~\ref{fig:jtime}. Hence, we argue that the formation path of individual galaxies is also reflected in the final
bulge properties. }

\section{The galaxy--halo connection}\label{sec:dm-baryon-connection}

\begin{figure*}
\vspace{-10.0pt}
\centering
\includegraphics[scale=.7]{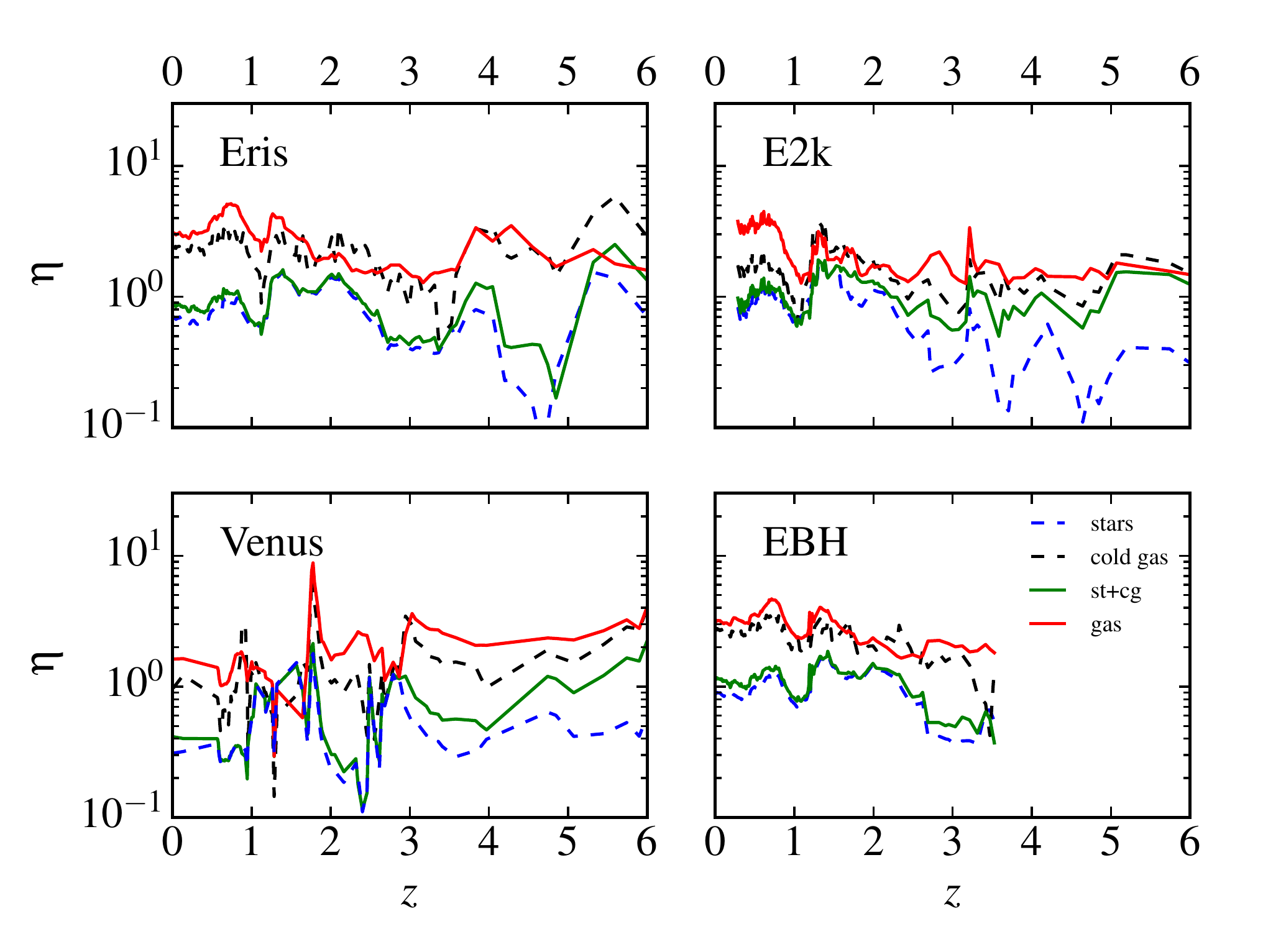}
\caption{Evolution of the retention factor of the different components of our four galaxies: stars, cold gas ($T < 10^4$~K), all gas, and stars with cold gas within the virial radius at a given redshift. Available data for EBH and E2k exist only for $z < 4$ and $z > 0.3$, respectively.}
\label{fig:eta}
\end{figure*}

{As discussed in the Introduction, 
theory and observations suggest that the specic angular momenta of galaxies, $j_{\rm galaxy}$, and those of their dark halos, $j_{\rm halo}$, are approximately proportional to each other. As the masses of galaxies and halos grow by accretion and merging, their specific angular momenta will also grow (on average). Thus, specific angular momentum is not strictly conserved even in the case $j_{\rm galaxy}=j_{\rm halo}$. We expect to learn more about the interplay between the baryons and dark matter in the galaxy formation process by examining the relations between the specific angular momenta of galaxies, that of their disk and bulge components, and that of their dark halos, as well
as how these relations evolve with redshift.}

{We compute the specific angular momenta of stellar disks, stellar bulges, gas at all temperatures, cold gas and dark matter. For each of these components, we define the angular momentum ``efficiency" or ``retention factor" $\eta\equiv j/j_{\rm halo}$. Unless otherwise stated, we include all the mass and angular momentum within the virial radius $R_{\rm vir}$ of the halo. It is worth noting that the components of galaxies, as we define them, do not consist of a fixed set of particles; at each redshift, particles are incorporated into or expelled from each component, depending on how galaxies evolve.}

{Figure~\ref{fig:jdm} shows the evolution of the specific angular momentum of} stars (blue dashed line), cold gas ($T<10^4$~K, black dashed line), total gas (red solid line), stars and cold gas (green solid line), and dark matter (black solid line) { in our four simulations}. Although the specific angular momentum of {each} component {generally} increases with time, {as expected, there are} noticeable fluctuations at nearly all redshifts. Those temporary gains and losses of the angular momentum per unit mass are stronger before $z=1$ in all the runs, regardless of the initial conditions. It has been already shown that the angular momentum can be both {decrease}, e.g. due to torques associated with violent disk instabilities \citep{Danovich:2015}, or increase in galactic fountains if material is ejected for long times and to large radii \citep{Ubler:2014}. 

\citet{Zavala:2016} report a better agreement between the specific angular momentum of the luminous matter and the dark matter within 10\% of the virial radius, rather than the full $R_{\rm vir}$, in their large-volume simulations. We investigate this possibility and show the specific angular momentum for the dark matter within $0.1 R_{\rm vir}$ with a solid grey line in Figure~\ref{fig:jdm}. On average, the specific angular momentum of stars evolves closer to that of dark matter within {$R_{\rm vir}$ rather than $0.1 R_{\rm vir}$}. This holds in all runs, except for Venus in two periods of time: between $z = 2.5$--2 and $z = 1$--0. Those two transitions follow sharp changes (reaching an order of magnitude) in the specific angular momentum of all components and are likely associated with major mergers. At those times, the $B/T$ ratios fluctuate (see Figure~\ref{fig:jtime}).

\begin{figure*}
\vspace{-10.0pt}
\hspace{-.7cm}
\centering
\includegraphics[scale=.7]{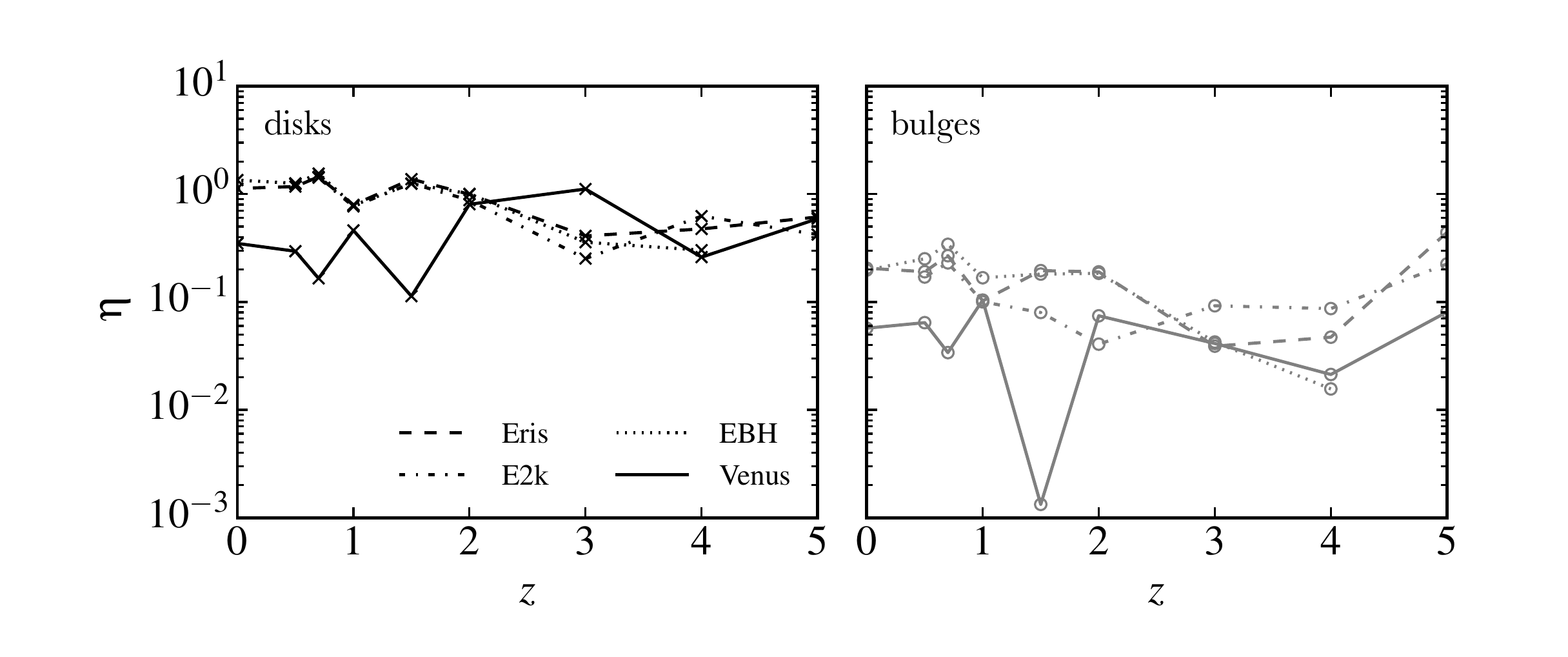}
\caption{Stellar retention factor for disks (left panel) and {bulge}s (right panel) calculated at a few redshifts, following the kinematic decomposition covered in Section~\ref{sec:j-M-diagrams}. Values for different components are marked with distinctive symbols: disks with crosses and {bulge}s with open circles). The lines denote the interpolations between these data points.}
\label{fig:etabd}
\end{figure*}

As pointed out in Section~\ref{sec:j-M-diagrams}, different initial conditions (active vs. quiet) translate into different locations of the evolutionary tracks of galaxies in Figure~\ref{fig:jtime}. As a result, Venus evolves closer to the {sequence of elliptical galaxies} than the other runs. Since the sequence of ellipticals is offset from that of disks by a factor of $\sim$5, the retention factor for ellipticals is expected to be lower at the same stellar or halo mass \citep[][]{Fall:2013}. {Assuming that the retention factor of pure disks is 0.8--1, one can infer from the $j_*$--$M_*$ diagram that a retention factor of a galaxy with the same properties as Venus ($M_*$ and $j_*$) is 0.2. This is close to our simulated value of 0.3. }

{Figure~\ref{fig:eta} shows the evolution of the retention factors of the baryonic components of our simulated galaxies. The color-coding is the same as in Figure~\ref{fig:jdm}. Evidently, the retention factors exhibit many fluctuations, but, overall, their average evolution is relatively mild.} For our sample of simulated galaxies, the stellar retention factor is $\eta \simeq (0.7, 0.8, 0.3, 0.9)$ at $z_{\rm end}$ for Eris, E2k, Venus, and EBH, respectively. {The t}otal gas and cold gas are endowed with a specific angular momentum that is 2--6 times higher than that of dark matter (the peak of Venus attains $\eta = 10$ in a major merger). Generally, the retention factor of the stellar component is confined within 0.1--2 in all cases{,} but most of the time below $z = 3$ it does not exceed 1.5 or fall below 0.5 in the runs with a quiet merger history. {The stellar retention factors in all runs are remarkably constant below $z = 1$, which agrees with the theoretical prediction $j\propto M^{2/3}$ for dark halos. }Interestingly, despite the relatively violent merger history, the retention factor of stars in Venus at $z=0$ is almost the same as {the initial value (at} $z = 6$). {Low-resolution, large-volume simulations have converged on the value of the retention factor $\eta$ for galactic disks that is approximately unity} \citep[][]{Genel:2015,Pedrosa:2015,Teklu:2015}. However, {\cite{Zavala:2016} found somewhat lower values of $\eta$ for disks, consistent with a better match to the dark matter within $0.1 R_{\rm vir}$ rather than $R_{\rm vir}$.}

Given the high resolution of our simulations, we can use the results of the last two sections -- careful decomposition and the calculation of the specific angular momentum of the stellar components -- to characterize the individual stellar retention factors of disks and {bulge}s at the time steps discussed in Section~\ref{sec:j-M-diagrams}. In Figure~\ref{fig:etabd}, we present for the first time the {evolution of the retention factors for disks and bulges separately over the redshift range $0 < z < 5$. The retention factors for disks are} remarkably constant, ranging from 0.3--1.5, over more than 12~Gyr of evolution. This means that the evolution of the rotationally supported component is driven by its dark matter halo, and this relationship is {tightest} for systems with quiet merger {histories. The retention factors for bulges also evolve relatively slowly but are much lower (by factors $\sim6$) than those for disks. Taken together, these results {indicate that galaxies and their disk and bulge components maintain nearly constant relationships} to their dark halos (after smoothing over short-term fluctuations). That is, galaxies and their halos evolve {\it quasi-homologously}.}

The results presented in this section raise questions about the physical mechanisms that affect galactic angular momentum, as well as the reasons for relatively constant proportionality between $j_{\rm disk}$, $j_{\rm bulge}$, and {$j_{\rm halo}$}. {We plan to address these questions in Paper~II. It appears that the strength of SN feedback or the presence of Milky Way--like AGN (thermal) feedback has less impact on the retention factor within $R_{\rm vir}$ than the merging history driven by the environment in which a galaxy is born \citep[see also the discussion in ][on how galaxy properties depend intimately on their environment]{Creasey:2015}. This conclusion, however, comes from a set of simulations employing only one type of a feedback model (blastwave feedback), thus calling for further investigation. }

\section{Conclusions}\label{sec:conclusions}

In this paper, we {present} an analysis of the angular momentum evolution of galaxies residing in Milky Way-sized halos, studying the relation between morphological appearance and kinematics as well as the evolutionary tracks on the $j_*$--$M_*$ diagrams. We use high-resolution cosmological zoom-in simulations, an approach that is essentially complementary to the large-volume calculations, as we are able to follow the assembly of each object separately and in greater detail. Our sample of simulations comprises runs with varying processes (SN and AGN feedback, as well as different radiative-cooling recipes), different strength of feedback, and different assembly histories for galaxies in halos of identical masses by $z = 0$.

{Thanks} to the high {spatial} resolution obtained in our simulations, we can also study the specific angular momentum evolution of the stellar components -- the disks and {bulge}s -- {as the gravitational softening length is an order of magnitude smaller than the characteristic sizes of these components.}  Additionally, a core part of this paper is devoted to studying the specific angular momentum of gas and dark matter. In what follows, we summarize the main findings of this work and motivate the necessity of a follow-up study.

\begin{enumerate}

\item The kinematic decomposition at $z = z_{\rm end}$ ($z_{\rm end} = 0$ for Eris, EBH, and Venus; $z_{\rm end} = 0.3$ for E2k), based on the circularity diagrams as a measure of the kinematics in the plane of the galaxy, and the photometric method yield results that are in perfect agreement for Eris and Venus, slightly deviated in the case of EBH and {are} off by a factor of 3 {for} E2k (see Section~\ref{sec:bulge-disk-decomposition}).
 
\item Our simulated galaxies {display} a variety of morphological types and lie on the $j_*$--$M_*$ diagrams with the population of spiral galaxies (Figure~\ref{fig:jstK}). When decomposed into disks and {bulge}s, the dichotomy in the specific angular momentum of disks and bulges is reproduced. The disks and bulges of our individual galaxies are separated by a factor 5.5--6.8, in agreement with the findings of {\cite{Fall:2013}}. Our galaxies do not suffer from the angular momentum problem and are good laboratories for the in-depth studies of the angular momentum evolution.

\item We present time-dependent diagrams (Figure~\ref{fig:diagrams}) that reveal correlations between the morphological appearance and the stellar kinematics of simulated galaxies, indicating that the latter can be predicted to some extent from the former.

\item {We inspected evolutionary tracks of individual galaxies on a physically-motivated equivalent of
the Hubble sequence ($j_*$--$M_*$ diagram).} On average, galaxies evolve on straight lines past major mergers on the {$j_*$--$M_*$} diagram (Figure~\ref{fig:jtime}). Eris and EBH evolve on {tracks with $\alpha=1.4$ (where $\alpha=d\log j_*/d\log M$).} This is likely due to the fact that they undergo a series of morphological changes, which in turn modify their $B/T$ ratios. E2k, which exhibits the least variations in this respect, evolves on {a track with} $\alpha = 2/3$. We argue that galaxies with relatively stable morphologies and secular processes occurring on long timescales move on the $\log j_*$--$\log M_*$ diagrams along these parallel {tracks}. Shorter timescale processes could perturb these tracks: frequent mergers may bring these galaxies closer to the tracks of ellipticals, as in the case of Venus. Although our sample of galaxies is too small to test these scaling relations at a fixed redshift, the time-dependent $j_*$--$M_*$ sequence of a single galaxy with a constant $B/T$ ratio provides an indirect test of this relationship. Recent {observational} results of \citet{Burkert:2016}, \citet{Contini:2016}, {and \cite{Huang:2016}} for disk-dominated galaxies at higher redshifts ($0.2 < z < 3$) lend support to this conjecture.

\item The specific angular momentum of baryons within $R_{\rm vir}$ tracks that of the halo (Figure~\ref{fig:eta}). The value for the total gas and cold gas is 2--6 times higher than that for the dark matter. The stellar retention factor is nearly constant below $z=1$ and reaches $(0.7, 0.8, 0.3, 0.9)$ at $z_{\rm end}$ in Eris, E2k, Venus, and EBH, respectively. {On average, the retention factors of baryonic components evolve weakly with redshift.}

\item In general, the specific angular momentum of stars is more consistent with that of the dark matter within the virial radius {$R_{\rm vir}$} rather than within {$0.1 R_{\rm vir}$ (}Figure~\ref{fig:jdm}). When a galaxy is disrupted by many major mergers (e.g. Venus), the overall specific angular momentum of stars is lowered, bringing the latter closer to the angular momentum content of a central subregion smaller than the virial region.

\item Galactic disks have nearly constant retention factors of order unity, which implies their {specific angular momenta are strongly correlated with those of the dark halos}. Exceptions are the phases in which major mergers occur, which is more relevant for Venus than for the other galaxies having quiet merging histories. The retention {factors for bulges are in general within a factor of 10 lower for our simulations and are also relatively constant when short-term fluctuations are smoothed out. These results lead to the notion that galaxies and their halos evolve quasi-homologously}.

\end{enumerate}

The good agreement with recent observations for angular momenta of low-redshift galaxies \citep{Fall:2013}, and the confirmation of a close connection of the specific angular momentum of dark matter and baryons, in particular of the disks, set the groundwork for an in-depth study of physical processes driving the evolution of the angular momentum of baryons in these galaxies. This will be presented in a follow-up paper (Paper~II). Paper~II will also investigate the role of feedback processes and disk instabilities in the angular momentum transport in these galaxies, as well as how the angular momentum of accreted gas evolves up to the point it joins the disk, shedding more light on the relation between the angular momentum of galaxies and that of their host halos.

\begin{comment}
studied the bulge in the Eris simulation, finding that the bulge was primarily originating from disk material that underwent repeated bar-like instabilities from high to low redshift, while (minor) mergers were found to suppress temporarily bulge growth by stirring the stars and reducing their central concentration due to SN-driven outflows following a central starburst. Although they argued that the bulge of Eris at low redshift is a pseudobulge, they based such conclusion mainly on the S\'{e}rsic index and colors, and did not consider the other properties used to classify bulges  which we tested in this paper.
\end{comment}

\section{Acknowledgements}
We thank Claudia Lagos, Shy Genel, and Tom Abel for helpful comments. This research was supported in part by the National Science Foundation under Grant No. PHY11-25915. LM and AS thank the Kavli Institute for Theoretical Physics for hospitality during the ``Cold Universe'' Program. PRC acknowledges support by the Tomalla Foundation.

%\bibliographystyle{aastex}
%\bibliography{apj-jour,references}
\bibliography{references}

\appendix

\section{Nature of bulges}\label{sec:spheroids}

As mentioned in Section~\ref{sec:bulge-disk-decomposition}, our sample of galaxies might contain classical bulges (C), pseudobulges (P), ``peanut'' bulges (box), or so-called composite bulges \citep[COMP, e.g.][]{KormendyBarentine:2010}. Here, we use six tests commonly used in literature (summarized in Table~\ref{tab:tab2}) in order to classify the {bulges} in our simulations, as well as qualify how sensitive the key results of this paper are to this categorization. These criteria are: visual morphology, presence of a bar, S\'{e}rsic index, size--mass relation, star formation rate and vertical distribution \citep[for more details, see e.g.][]{Gadotti_DosAnjos_2001, Kormendy:2004, Gadotti:2009}.

Based on pure visual appearance when seen edge-on (see the last row in Figure~\ref{fig:diagrams}), the bulges of Venus and Eris could be classified as classical bulges at $z=0$, whereas those of E2k and EBH appear {flatter and more} disky, hence more similar to pseudobulges.

The fact that EBH and E2k are the only two galaxies that host a strong, large-scale bar at low redshift {reinforces this distinction} (Eris has a bar at higher redshift which weakens and shortens at low redshift, becoming essentially a nuclear bar a few gravitational softenings long).

In terms of the S\'{e}rsic index, all galaxies have a rather low $n$ of order 0.8--1.4 at $z_{\rm end}$ (see Table~\ref{tab:tab1}). Given that a relatively low S\'{e}rsic index photometrically akin to a disk component ($n < 1.5$) is characteristic of pseudobulges, the values obtained from the photometric decomposition are hardly indicative of classical bulges.

\begin{figure}
\centering
\includegraphics[scale=0.65]{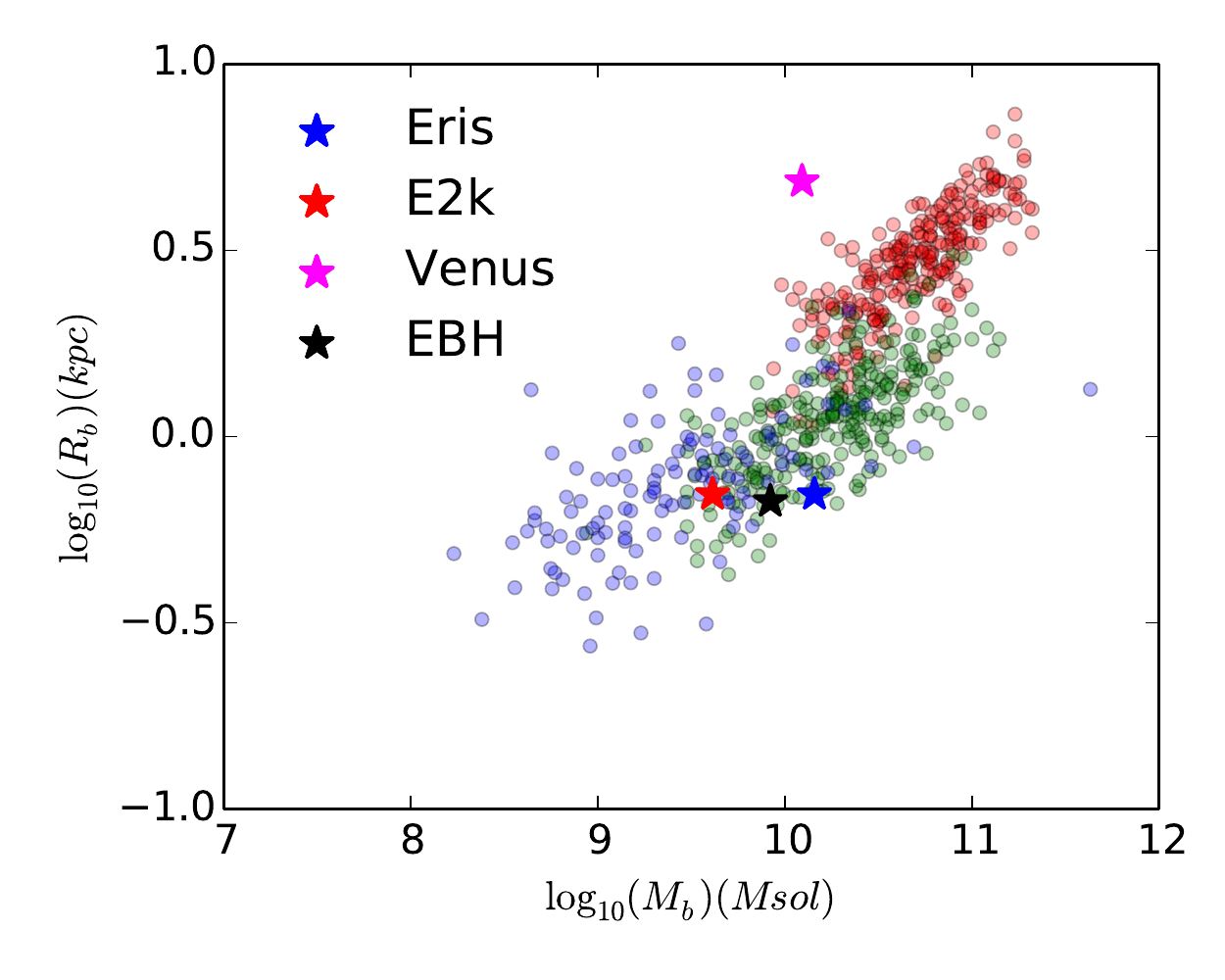}
\caption{The comparison of bulge scale lengths of our sample of galaxies after the photometric decomposition as a function of the stellar mass of their {bulges} (stars) vs. the sample of SDSS elliptical galaxies (red circles), classical bulges (green circles), and pseudobulges (blue circles) from \citet{Gadotti:2009}.}
\label{fig:re_size}
\end{figure}

Another criterion utilizing our results of the photometric decomposition is the mass--size relation. In Figure~\ref{fig:re_size}, we compare the location of our bulges on the mass--size diagram with the sample of \citet{Gadotti:2009}, who found unique relations for ellipticals, classical bulges, and pseudobulges. Our galaxies lie on that diagram in the sequence of decreasing importance of the bar with increasing mass, which places E2k in the area of the graph populated by pseudobulges, Eris by classical bulges, and EBH at the intersection of the two. We note that Figure~\ref{fig:re_size} also places Venus high above the line of ellipticals due to a large radius $R_{\rm b}$, which was already argued in Section~\ref{sec:bulge-disk-decomposition} to be unrealistic and likely represents a failure of the photometric method of decomposition. A brighter stellar envelope of Venus suggests that a more prominent stellar halo or a thick disk component contaminates the decomposition. Indeed, restricting the region to a slice of height 1~kpc above and below the disk plane reduces the bulge scale length to $\sim$3~kpc.

\begin{figure*}
\centering
\includegraphics[scale=0.8]{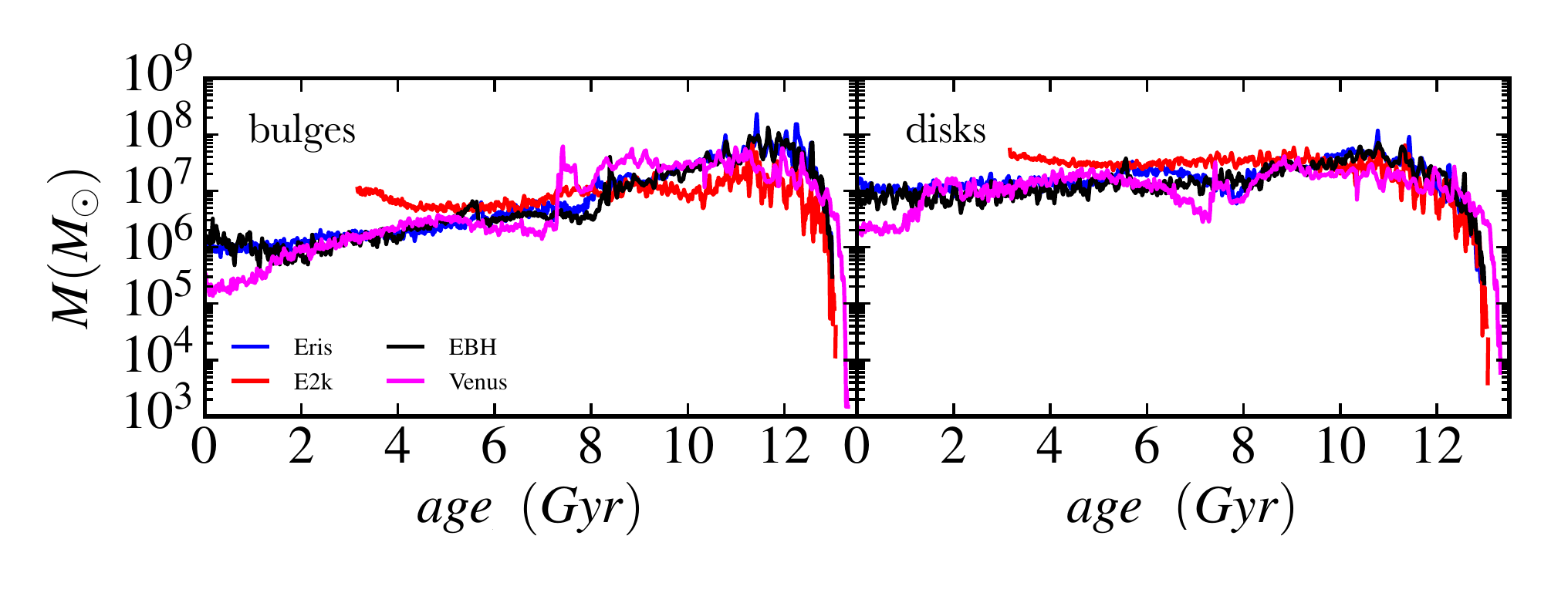}
\caption{Mass distribution of stellar ages of {bulge}s (left) and disks (right), as would be measured at $z=0$. Note that the lack of stars of ages lower than 3.5~Gyrs in E2k is due to $z_{\rm end}=0.3$ of that run. }
\label{fig:age}
\end{figure*}

The fifth criterion, stellar ages, is addressed in Figure~\ref{fig:age}, where we show the mass distribution of stellar ages of bulges (left) and disks (right). A massive population of young stars (younger than 4~Gyr) in the bulge of E2k is characteristic of still star-forming pseudobulges. At another extreme, Venus experiences a sharp decline in this distribution near 2~Gyr at the level clearly indicating quenching of star formation, which would be consistent with what is expected of a classical bulge. Eris, EBH, and Venus are good examples of galaxies with a rather old bulge and a young disk.

\begin{figure}
\centering
\includegraphics[scale=0.5]{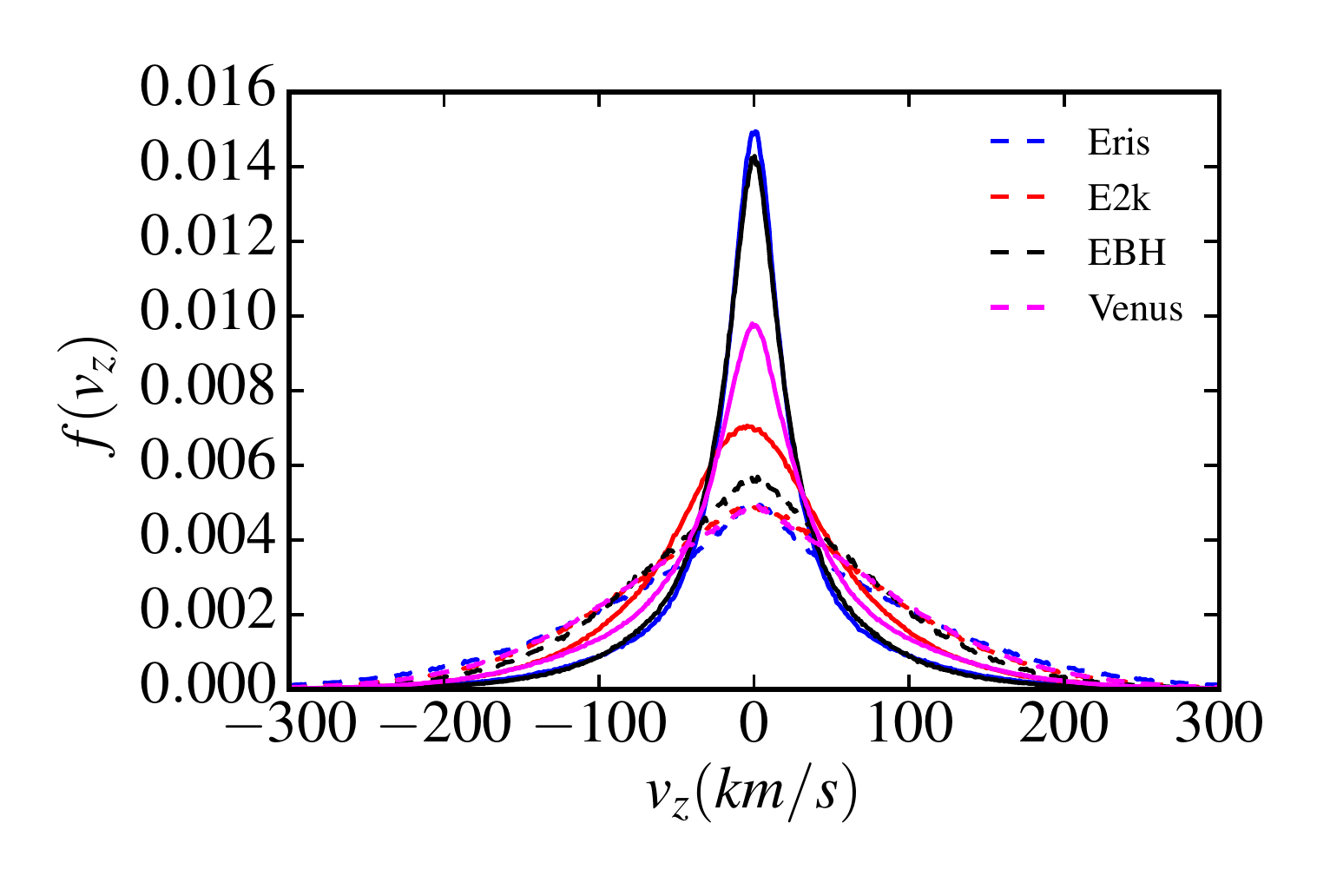}
\includegraphics[scale=0.5]{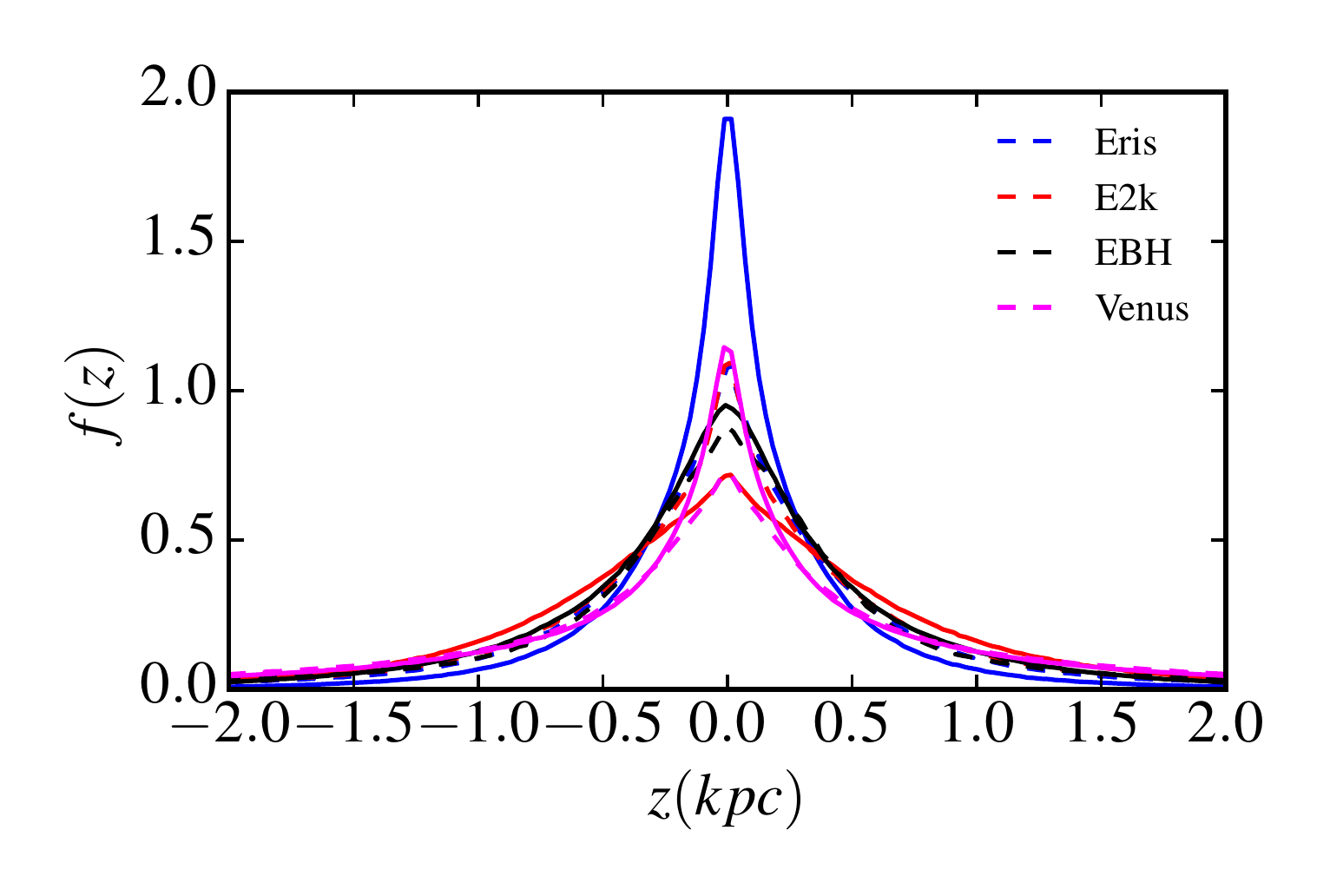}
\caption{Distribution of the vertical kinematics and vertical locations of the stellar particles in galactic bulges (dashed lines) and disks (solid lines) at $z=z_{\rm end}$.}
\label{fig:disp}
\end{figure}

The final criterion, the vertical kinematics, allows for determining whether disks and {bulge}s are kinematically and structurally alike. Hence, in the top and bottom panels of Figure~\ref{fig:disp}, we show the distribution of the vertical velocity and vertical position of the stellar particles, respectively. The components of Eris and EBH have very distinct vertical kinematics, which is an attribute of galaxies with classical bulges. In contrast, E2k exhibits very little distinction, whereas the case of Venus is rather ambiguous, yet closer to the classical picture. In the bottom panel of Figure~\ref{fig:disp}, the vertical position distributions of the bulge and disk particles of EBH are nearly identical, whereas the disk of E2k is thicker than the {bulge}, both certainly indicative of a pseudobulge. The bulges of Eris and Venus have clearly broader distributions than the disks, hence a significant part of their mass is off-planar.

%\tabletypesize{\footnotesize}
\begin{deluxetable}{l|c|c|c|c}
\tablecaption{Types of bulges in the simulations at $z_{\rm end}$}
\tablenum{3}
\tablehead{\colhead{Criterion} & \colhead{Eris} & \colhead{Venus} & \colhead{E2k} & \colhead{EBH}}
\startdata
morphology &C &C & P & box \\
bar & P& C& P&P \\
S\'{e}rsic & P&P &P &P \\
size--mass & C&C &P &C/P \\
star formation & C/P& C&P &C/P \\
vertical distribution & C& C&P &C/P \\
\hline
summary & COMP& C&P & box\\
\enddata
\tablecomments{The results for each classification scheme are labelled as: P -- pseudobulge, C -- classical bulge, box -- peanut bulge, COMP -- composite bulge. The ``morphology'' criterion is based on the surface density stellar maps.}
\label{tab:tab2}
\end{deluxetable}

Although the criteria do not always agree on the classification, overall there is more evidence that {E2k and EBH have} pseudobulges, whereas {Venus has} a classical bulge. The bulge of Eris, however, appears to be a composite bulge, i.e. a small, star-forming disk-like bulge inside a classical bulge (see the last row of Table~\ref{tab:tab2}). Our conclusion categorizing the bulge of Eris as a composite bulge rather than a pseudobulge complements the previous in-depth studies of the bar evolution in that simulation \citep{Guedes:2013}.

Interestingly, if we look at the problem from the formation point of view, we can notice an interesting trend in Figure~\ref{fig:jtime}{, namely that the most classical-like bulge, that of Venus, evolves most differently from its disk, while the most pseudo-like bulge evolves the most similarly to its disk.} Hence, our results for {bulge}s shown in Figure 5 verify that classical bulges (Venus) have lower angular momentum than pseudobulges or composite bulges (the remainder). Also, while in Venus there is a clearly different evolutionary track between the stars as a whole and the bulge, these tracks are almost coincident in the other three cases. This reinforces the notion that formation {of bulges} is reflected in {their} final bulge properties, although there is enough diversity and scatter in such properties, and in the results of different diagnostics, to preclude any rigorous statements.

\end{document}